\documentclass[article,twocolumn,final,10pt]{IEEEtran}

\usepackage{amsmath}
\usepackage{amsfonts}
\usepackage{dsfont}
\usepackage{mathrsfs}
\usepackage{pifont}
\usepackage{amssymb}
\usepackage{verbatim}
\usepackage{upgreek}
\usepackage{color}
\usepackage[final]{graphicx}
\usepackage{graphicx}
\usepackage{caption}
\usepackage{algorithmic}
\usepackage{algorithm}
\usepackage{epsfig}
\usepackage{eucal}
\usepackage{cite}
\usepackage{subfigure}

\newcommand{\bbm}{\begin{bmatrix}}
\newcommand{\ebm}{\end{bmatrix}}

\newcommand{\bit}{\begin{itemize}}
\newcommand{\eit}{\end{itemize}}

\newcommand{\ben}{\begin{enumerate}}
\newcommand{\een}{\end{enumerate}}

\newcommand{\bdesc}{\begin{description}}
\newcommand{\edesc}{\end{description}}

\newcommand{\bea}{\begin{array}}
\newcommand{\eea}{\end{array}}

\newcommand{\tr}{\mbox{\rm Tr}\, }

\newcommand{\beqa}{\begin{eqnarray}}
\newcommand{\eeqa}{\end{eqnarray}}

\newcommand{\Comment}[1]{}

\def\R{{\mathds R}}

\def\C{{\mathds C}}

\newcommand{\be}{\begin{equation}}
\newcommand{\ee}{\end{equation}}

\newcommand{\bzero}{{\mbox{\boldmath $0$}}}

\newcommand{\boa}{{\mbox{\boldmath $a$}}}

\newcommand{\bn}{{\mbox{\boldmath $n$}}}

\newcommand{\bv}{{\mbox{\boldmath $v$}}}
\newcommand{\bx}{{\mbox{\boldmath $x$}}}

\newcommand{\bz}{{\mbox{\boldmath $z$}}}

\newcommand{\bI}{{\mbox{\boldmath $I$}}}

\newcommand{\bP}{{\mbox{\boldmath $P$}}}

\newcommand{\bR}{{\mbox{\boldmath $R$}}}
\newcommand{\hbR}{{\mbox{$\widehat{\bR}$}}}
\newcommand{\bS}{{\mbox{\boldmath $S$}}}

\newcommand{\bT}{{\mbox{\boldmath $T$}}}

\newcommand{\bV}{{\mbox{\boldmath $V$}}}

\newcommand{\diag}{\mbox{\boldmath\bf diag}\, }

\newcommand{\balpha}{{\mbox{\boldmath $\alpha$}}}

\newcommand{\dmax}{\begin{displaystyle}\max\end{displaystyle}}

\newcommand{\test}{\mbox{$
\begin{array}{c}
\stackrel{ \stackrel{\textstyle H_1}{\textstyle >} }{
\stackrel{\textstyle <}{\textstyle H_0} }
\end{array}
$}}

\makeatletter
\def\maketag@@@#1{\hbox{\m@th\normalfont\normalsize#1}}
\makeatother

\title{A Sparse Learning Approach to the Design of Radar Tunable Architectures with Enhanced Selectivity Properties}

\vspace{0.1cm}

\begin{document}
\author{Sudan~Han,
		Luca~Pallotta,~\IEEEmembership{Senior Member,~IEEE,}
		Xiaotao~Huang,~\IEEEmembership{Member,~IEEE,}
		Gaetano~Giunta,~\IEEEmembership{Senior Member,~IEEE,}
		and~Danilo~Orlando,~\IEEEmembership{Senior Member,~IEEE}
		\thanks{S. Han is with the National Innovation Institute of Defense Technology, Beijing, China E-mail: xiaoxiaosu0626@163.com.}	
		\thanks{L. Pallotta and G. Giunta are with Department of Engineering, University of Roma Tre, via Vito Volterra 62, 00146 Rome, Italy. E-mail: {luca.pallotta@uniroma3.it, gaetano.giunta@uniroma3.it.}}
		\thanks{X. Huang is with the College of Electronic Science and Technology, National University of Defense Technology, Changsha 410073, China. E-mail: xthuang@nudt.edu.cn.}
		\thanks{D. Orlando is with Universit\`a degli Studi ``Niccol\`o Cusano'', via Don Carlo Gnocchi 3, 00166 Roma, Italy. E-mail: {danilo.orlando@unicusano.it}.}}

\maketitle
	
\begin{abstract}
This paper considers the design of tunable decision schemes capable of rejecting with high probability 
mismatched signals embedded in Gaussian interference with unknown covariance matrix. 
To this end, a sparse recovery technique is exploited to enhance the resolution at 
which the target angle of arrival is estimated with the objective to obtain high-selective detectors. 
The outcomes of this estimation procedure are used to devise detection architectures relying on 
either the two-stage design paradigm or heuristic design procedures based upon the 
generalized likelihood ratio test. 
Remarkably, the new decision rules exhibit a bounded-constant false alarm rate property and allow for 
a tradeoff between the matched detection performance and the rejection of undesired signals by 
tuning a design parameter. At the analysis stage, the performance of the newly proposed detectors 
is assessed also in comparison with existing selective competitors. The results show that 
the new detectors can outperform the considered counterparts in terms of rejection of 
unwanted signals, while retaining reasonable detection performance of matched signals.
\end{abstract}


\section{Introduction}
In the recent years, the design of adaptive detection architectures in the presence of mismatched signals has raised a strong interest in the radar community as corroborated by the multitude of contributions that can be found in the open literature \cite{kalson1992mismatched, bandiera2009, hao2010, liu2014, bandiera2008subspaceasb, bandiera2007conic, bandiera2009conic, bandiera2008asb, pulsone2000, richmond2000asb, richmond1997asb, hao2011, demaio2016twostage, danilobook, bandiera2008abort, liu2017, demaio2005mismatched, pulsone2001about,demaiobook,InterferenzaCono}. 
As a matter of fact, in scenarios of practical interest, the direction of arrival of the signal backscattered 
from a target may be different from the nominal pointing direction of the mainbeam due to environmental and/or instrumental factors. For instance, in search mode, radar operates in situations where a possible 
target echoes could come from any arbitrary angle within the beamwidth with a consequent loss in detection performance and a biased estimate of the target direction. 
On the other hand, the presence of a coherent jammer or a strong target in the sidelobes (namely, 
from a significantly mismatched direction of arrival) could trigger a detection which deceives 
the search system. Other causes of mismatched signals may be imperfect modeling of the mainlobe steering vector, multipath propagation, array calibration uncertainties, mutual coupling between the array elements, etc. 

Conventional adaptive detection algorithms, namely those designed under the assumption of a perfect match 
between the nominal and the actual steering vector \cite{kelly1986adaptive, robey1992cfar, demaio2007rao, orlando2010rao, liu2016rao, demaio2016symmetric, liu2015persymmetric}, behave quite differently in the presence of mismatched signals. Specifically, according to their directivity, defined as the capability of rejecting/detecting mismatched signals, they can be classified as \cite{demaio2016twostage}:
\begin{itemize}
\item robust decision schemes which provide good detection performances in the presence of echoes containing 
signal components not aligned with the nominal (transmitted) signal (see, for instance, the subspace detectors 
\cite{DD,SD,scharfSD});
\item selective decision schemes which are capable of rejecting signals whose signature is unlikely 
to correspond to that of interest in order to avoid false alarms (see, for instance, 
\cite{bandiera2008abort,pulsone2001about,scharf1999ace,orlando2010rao}).
\end{itemize}

From an operating point of view, selective detectors may be exploited to face with densely populated environments 
or to face with electronic countermeasures (coherent jammers) \cite{EW101}. On the other hand, robust architectures are suitable to 
cover wide angular sectors by means of a low number of filters/pointing directions or to detect 
mismatched signals within the mainbeam. 
As corroborated by the numerous analysis in the presence of mismatched signals, generally speaking, 
an enhanced selectivity degrades the matched detection performance, whereas robust architectures
can maintain good matched detection performance \cite[and references therein]{danilobook}.
Thus, it is clear that a decision scheme capable of modifying its behavior according to the specific scenario would 
offer a relevant flexibility in usage. This requirement has led to the birth of tunable detectors whose 
directivity can be set by means of suitable design parameters. 
Remarkably, they are capable of providing a good tradeoff between matched detection performance 
and rejection of unwanted signals.

There exist several design paradigms to come up with tunable architectures. A first approach consists in merging 
the decision statistics of existing detectors due to the inherent similarities between 
them \cite{kalson1992mismatched, 
bandiera2009, hao2010, liu2014}. As a result, the new architecture encompasses the merged 
decision schemes as special cases by tuning suitable design parameters.
Another class of tunable architectures can be formed by resorting to the theory of subspace detection. 
Specifically, at the design stage, it is assumed that the possible useful signals belong to a preassigned 
subspace of the observables or to a proper cone with the nominal steering vector as axis 
\cite{bandiera2008subspaceasb, bandiera2007conic, bandiera2009conic}. In these cases, the directivity
can be modified by acting on the subspace properties or on the cone aperture.
Finally, a powerful tool for the design of tunable architectures is the two-stage approach, which 
consists in cascading two decision schemes (usually with opposite behaviors in terms of directivity) 
\cite{bandiera2008asb, pulsone2000, richmond2000asb, richmond1997asb, bandiera2009, hao2011, demaio2016twostage}.
The overall detector decides for the alternative hypothesis if and only if the decision statistic of each stage is above the respective threshold. This scheme is tantamount to a logical AND between the two stages and it can be tuned by modifying the two thresholds. In fact, a preassigned value for the probability of false alarm ($P_{fa}$) can be maintained over different threshold pairs and, interestingly, each pair returns a specific directivity
and matched detection performance \cite{danilobook,richmond2000asb,pulsone2000}.

In this paper, we focus on the design of tunable adaptive architectures capable of achieving an enhanced
selectivity with respect to the state-of-the-art selective detectors while keeping good detection
performance for matched signals embedded in Gaussian interference with unknown covariance matrix. 
To this end, we exploit
sparse reconstruction techniques in conjunction with either the two-stage design paradigm or 
{\em ad hoc} modifications of the generalized likelihood ratio test (GLRT) \cite{robey1992cfar}.
More precisely, at the design stage, we take advantage of the inherent sparse nature of the signal model due 
to the fact that a large part of the angular sector under surveillance does not contain any signal 
of interest leading to a sparse scene. Therefore, we resort to the compressive sensing approach to estimate target response as well as its angle of arrival (AOA) \cite{donoho2006cs, Tropp2007OMP, donoho2006, stoicaSparse, rossi2014}
and use such estimates to build up two new classes of decision schemes. 
The first class is represented by two-stage architectures obtained by coupling the sparse amplitude detector (SAD), 
defined in Section \ref{Sec:decision scheme}, and classical constant false alarm rate (CFAR) detectors 
such as Kelly's GLRT \cite{kelly1986adaptive} or the adaptive matched filter (AMF) \cite{robey1992cfar}. 
The second class of receivers is devised applying a heuristic procedure based upon the GLRT, where
some parameters are assumed known and the others are estimated through the maximum likelihood approach
possibly resorting to training samples collected in the proximity of the cell under test.
Remarkably, it can be shown that the proposed architectures are bounded-CFAR. 

It is important to notice that
conversely to existing contributions on AOA estimation by means of sparse recovery, which focus on
estimation aspects only and assume that the estimation algorithms are triggered by a preliminary 
detection stage \cite{Malioutov2005, Hyder2010, Tan2014, liu_doa_2012,zhangCognitive,Sen}, 
the newly proposed architectures jointly perform detection and estimation.
In addition, the present approach allows us to tune the directivity of the proposed architectures 
through a design parameter
used by the sparse estimation procedure leading to enhanced rejection capabilities of unwanted signals.
The above issues appear for the first time (at least to the best of the authors' knowledge)
in this paper and represent the main technical contribution.

The performance analysis is conducted over simulated data by comparing the proposed techniques with 
Kelly's GLRT (which is considered as benchmark detector for matched signals), the AMF, 
and well-known selective architectures as the whitened adaptive beamformer orthogonal rejection test (W-ABORT) 
\cite{bandiera2008abort}, the Rao test (RAO) \cite{demaio2007rao}, and the adaptive coherence estimator (ACE)
\cite{scharf1999ace} (also known as the adaptive normalized matched filter (ANMF) \cite{conte1995ace}). 
The illustrative examples point out that, at least for the considered simulation parameters, 
the newly proposed architectures are capable of exhibiting an increased selectivity with respect to the considered
competitors and, at the same time, an excellent matched detection performance.

The remainder of the paper is organized as follows: Section \ref{Sec:Problem_Formulation} is devoted 
to the problem formulation. Section \ref{Sec:MAP+BIC algorithm} describes the sparse recovery procedure. 
The proposed decision schemes based on the sparse estimates are devised in Section \ref{Sec:decision scheme}, 
while the performance of the new detectors is assessed in Section \ref{performance assessment}. Finally, concluding remarks and future research tracks are given in Section \ref{conclusion}.

\subsection{Notation} 
Vectors and matrices are denoted by boldface lower-case and upper-case letters, respectively. 
The symbols $\det(\cdot)$, $\tr(\cdot)$, $(\cdot)^T$, and $(\cdot)^\dag$ denote the determinant, trace, transpose, 
and conjugate transpose, respectively. As to numerical sets, $\R$ is the set of real numbers, $\R^{N\times M}$ 
is the Euclidean space of $(N\times M)$-dimensional real matrices (or vectors if $M=1$), $\C$ is the set 
of complex numbers, and $\C^{N\times M}$ is the Euclidean space of $(N\times M)$-dimensional complex matrices 
(or vectors if $M=1$). The Euclidean norm of a generic vector $\bx$ is denoted by $\|\bx\|$. $\bI_N$ stands for the $N \times N$ identity matrix, while $\bzero$ is the null vector or matrix of proper size. Finally, given a vector $\boa$, $\diag(\boa)$ indicates the diagonal matrix whose $i$th diagonal element is the $i$th entry of $\boa$.

\section{Problem Formulation and Motivation}
\label{Sec:Problem_Formulation}

Let us consider a search radar system which exploits a uniform linear array with $N$ spatial channels and 
illuminates with its beam a given azimuth direction. The radar collects data from multiple range cells and 
tests whether or not the returns (from a specific range bin) contain a mainbeam target. In the positive case, 
it provides the AOA and the range measurement of the detected target. 
The classical detection procedure is implemented by testing range cell by range cell and it is often assumed that the actual target AOA coincides with the nominal steering angle (namely, the steering vector corresponding to the boresight). 
In this case, the detection problem at hand for a specific range bin, whose returns are collected
in the vector $\bz\in\C^{N\times 1}$, can be formulated in terms of the following hypothesis test
\be
\begin{cases}
H_1: \ \bz=\alpha\bv(\theta_p) + \bn,
\\
H_0: \ \bz=\bn,
\end{cases}
\label{eqn:nominalDetProb0}
\ee
where $\alpha\in\C$ accounts for transmitting antenna gain, the 
two-way path loss, and radar cross-section of the (slowly-fluctuating) target, 
$\bn\in\C^{N\times 1}$ represents the interference (clutter plus noise) component, modeled as a circular, zero-mean, complex Gaussian random vector with unknown positive-definite 
covariance matrix $\bR\in\C^{N\times N}$, $\theta_p$ is the nominal AOA of the target (which coincides with the beampointing direction), and $\bv(\theta_p)=\left[1, e^{j 2\pi (d/\lambda) \sin(\theta_p)}, \ldots, e^{j (N-1) 2\pi (d/\lambda) \sin(\theta_p)}\right]^T \in\C^{N\times 1}$ is the nominal (spatial) steering vector with $d$ the interelement spacing and $\lambda$ the operating wavelength. 
When $H_1$ is declared, the range associated with $\bz$ and $\theta_p$ are returned as target parameter estimates.
However, as mentioned in the previous section, in practice there exist several
factors that would make the perfect match assumption between the signature of the received echoes
and the nominal steering vector no longer valid. As a consequence, instead of problem (1), it would
be more plausible to consider the following alternative
\be
\begin{cases}
H_1 : & \bz= \alpha \bv(\theta_t) + \bn,
\\
H_0 : & \bz= \bn,
\end{cases}
\ee
where $\theta_t$ is the unknown AOA of the structured returns generated by an object 
in the surveillance area which 
may be different from the nominal pointing direction $\theta_p$. Notice that such model is more general than (1) 
since it accounts for situations where the structured component of the collected vectors 
may be generated by noninterest objects as,
for instance, a signal transmitted by a coherent jammer and entering from the sidelobes to inject false
information into the radar processor \cite{antennaBased,Farina-Handbook,giniGrecoDRFM}. 
Another example is represented by target-rich environments where an object within 
the $3$ dB mainbeam but not aligned with the antenna boresight may trigger a detection \cite{guerciMultTarget}.
In these operating scenarios, it would be desirable that the radar processor does not declare
$H_1$ (see also antenna-based sidelobe blanking techniques \cite{antennaBased,YanAddabbo,danilobook}). 
Otherwise stated, the decision rules incorporated in the radar processor should be
more inclined to decide for $H_0$ when the AOA of the coherent component is different from the nominal pointing
direction on the basis of a mismatch degree decided by the user according to the specific application 
and the system operating requirements.

A viable strategy to cope with the above situations could be a sequential test of different 
mainbeam azimuth positions. In this context, assuming the availability of 
a secondary data set $\bz_k \in \C^{N\times 1}$, $k=1,\ldots,K$, free of useful signal components but 
sharing the same spectral properties as the interference in $\bz$ (homogeneous environment), some 
classic decision rules can be used such as Kelly's GLRT \cite{kelly1986adaptive} and the AMF \cite{robey1992cfar}, 
which ensure excellent matched detection performance. Specifically, for an angular position $\theta$, 
the decision schemes for Kelly's GLRT and the AMF are given by
\be
\Lambda_{\mbox{\tiny GLRT}}=\frac{|\bv^\dag(\theta)\hbR^{-1}\bz|^2}{\left(\bv^\dag(\theta)\hbR^{-1}\bv(\theta)\right)\left(K+\bz^\dag\hbR^{-1}\bz\right)} \test \eta_{\mbox{\tiny GLRT}}, 
\ee
\be
\Lambda_{\mbox{\tiny AMF}}=\frac{|\bv^\dag(\theta)\hbR^{-1}\bz|^2}{\bv^\dag(\theta)\hbR^{-1}\bv(\theta)}
\test \eta_{\mbox{\tiny AMF}},
\ee
where $\hbR=\frac{1}{K}\sum_{k=1}^{K}\bz_k\bz_k^\dag$ is the sample covariance matrix (SCM) over the training data, 
$\eta_{\mbox{\tiny GLRT}}$ and $\eta_{\mbox{\tiny AMF}}$ are the thresholds set in order to ensure a given $P_{fa}$. 
The relationships between the thresholds and the given $P_{fa}$ are given by
\be
\mbox{Kelly's GLRT: \;} \eta_{\mbox{\tiny GLRT}}=1-P_{fa}^{\frac{1}{K-N+1}};
\label{eqn:Kelly_thr}
\ee
\be
\mbox{AMF: \;}
P_{fa}=\int_{0}^{1}f_{\beta}(\rho;L+1,N-1)\left(1+\eta_{\mbox{\tiny AMF}}\rho/K\right)^{-L}d\rho,
\label{eqn:AMF_thr}
\ee
where $L=K-N+1$ and
\be
f_{\beta}(x;n,m)=\frac{(n+m-1)!}{(n-1)!(m-1)!}x^{n-1}(1-x)^{m-1}
\ee
is the complex central Beta probability density function (pdf) \cite{kelly1986adaptive, robey1992cfar}.
Both Kelly's GLRT and the AMF ensure the CFAR property with respect to the interference covariance matrix. 
However, even though the performance in the case of matched signals is excellent, their azimuth discrimination as well as their capability of rejecting mismatched signals is limited. 
As a matter of fact, a target from directions that are different from the nominal pointing direction might trigger multiple detections (see \cite{danilobook} and references therein).

In order to circumvent this drawback, we devise two new classes of tunable architectures which take advantage
of sparse reconstruction techniques to achieve an enhanced selectivity.
More precisely, let us partition the angular region including the antenna mainbeam plus the relevant sidelobes 
into $M$ uniformly spaced azimuth bins with separation $\Delta\theta$. Now, denoting 
by $\theta_{l}$, $l=1,\ldots,M,$ the angle corresponding to the center of the $l$th azimuth bin, we model
the returns from a given range cell as follows
\be
\bz=\sum_{l=1}^{M}\alpha_{l}\bv(\theta_l)+\bn=\bV\balpha+\bn,
\label{eqn:z}
\ee
$\bV=[\bv(\theta_1),\cdots,\bv(\theta_M)]\in \C^{N\times M}$ denotes the (so-called) dictionary matrix and 
$\balpha=[\alpha_1,\cdots,\alpha_M]^T\in \C^{M\times 1}$ is the vector whose entries are the 
responses of prospective targets. 

Two remarks are now in order. First, note that
$\balpha$ is a sparse vector where the nonzero entry is that corresponding to the actual AOA 
of the target, whereas the other components are zero. Thus, \eqref{eqn:z} highlights the inherent 
sparse nature of the model under $H_1$ which allows to apply sparse reconstruction 
techniques \cite{donoho2006cs, Tropp2007OMP, donoho2006, stoicaSparse, rossi2014, sudan} 
to estimate $\balpha$. Finally, as shown in what follows, it is important to underline 
that in this context, $\Delta \theta$ can be 
viewed as a tuning parameter by which it is possible to control the angular estimation resolution 
and the estimation quality. As a matter of fact,
given $\Delta\theta$ ($N$), high values of $N$ ($\Delta \theta$) 
allow for a low dictionary coherence leading to high quality estimation for the sparse recovery technique.
In addition, under the above conditions, the inner product between 
adjacent columns of $\bV$ decreases and, consequently, the spillover of target energy
between consecutive azimuth bins takes on low values. However, high values of $\Delta \theta$ decrease
the angular resolution for the AOA estimation.
Thus, establishing the range of values for $\Delta \theta$ is not an easy task and requires a preliminary
analysis to find a good compromise between estimation resolution and quality also accounting for 
the specific application and the system operating requirements.

In the next section, we describe the sparse recovery algorithm used 
to estimate $\balpha$ assuming that secondary data are available for the estimation 
of the interference covariance matrix and that $N<M$ in order to obtain an overdetermined model. Then, the above estimates are suitably exploited in Section \ref{Sec:decision scheme} to build up adaptive detection architectures with enhanced rejection capabilities of unwanted signals.

\section{User Parameter Free BIC Based SLIM (BSLIM) Algorithm}
\label{Sec:MAP+BIC algorithm}
This section provides the description of the specific sparse recovery algorithm used to estimate $\balpha$. Such
algorithm relies on the sparse learning via iterative minimization (SLIM) method of \cite{stoicaSparse}. 
This design choice is dictated by the fact that SLIM approach exhibits a computational cost similar to 
the widely used compressive sampling matching pursuit (CoSaMP) \cite{NEEDELL2009301} but more accurate 
estimates than the latter, the iterative adaptive approach (IAA) \cite{5417172}, and matched filter techniques as 
shown in \cite{stoicaSparse}. Moreover, the interested reader is referred to 
\cite{liu2012correction,xu2013wideband,jabbarian2015two,addabbo2018hrr,feng20182,aubry2019multi} for
further applications/extensions of the SLIM.

In order to apply the SLIM method, we enforce a sparsity constraint on $\balpha$ using the same sparsity 
promoting prior pdf as in \cite{stoicaSparse}:
\be
f(\balpha)=\frac{1}{C} \prod_{l=1}^{M}\exp\left\{ -\frac{2}{q} \left(|\alpha_{l}|^q-1\right) \right\}, \quad 0<q \leq 1,
\label{eqn:prior_alpha}
\ee
where $C$ is a normalization constant which, without loss of generality, will be neglected in the sequel and $q$ is a user parameter controlling the sparsity of $\balpha$. In general, smaller $q$ leads to sparser estimates in the framework of Bayesian inference \cite{stoicaSparse}. The SLIM algorithm is based on a maximum a posteriori (MAP) approach, and thus given $\bz$ and $\bR$, the estimate of $\balpha$ can be written as
\be
\widehat{\balpha}=\arg\max\limits_{\balpha} f(\bz|\balpha;\bR) f(\balpha),
\label{eqn: map_alpha}
\ee
where
\be
f(\bz|\balpha;\bR)= \frac{1}{\pi^{M}\det(\bR)}\exp\left\{ 
-\left\| \bR^{-1/2}\left(\bz-\bV\balpha\right) \right\|^2
\right\}
\ee
is the conditional pdf of $\bz$ given $\balpha$. After some algebra, it is possible to show that
\eqref{eqn: map_alpha} is tantamount to
\be
\widehat{\balpha}=\arg\min\limits_{\balpha} g_q(\balpha,\bR),
\label{eqn: slim_objective_modify}
\ee
where
\be
g_q(\balpha,\bR)=\left\| \bR^{-1/2}\left(\bz-\bV\balpha\right) \right\|^2 
+ \sum_{l=1}^{M} \frac{2}{q} \left(|\alpha_{l}|^q-1\right).
\label{eqn: slim_objective}
\ee

Before proceeding towards the solution of the above problem, we focus on the interference covariance matrix $\bR$, 
that in practice is unknown. For this reason, the radar system
collects training samples in proximity of that under test \cite{Richards} and that are representative
of the interference affecting the cell under test. Thus, in what follows, we estimate $\bR$ by 
means of the SCM based on secondary data, namely $\hbR$.
As a consequence, after replacing $\bR$ with the considered estimate, problem \eqref{eqn: slim_objective} 
becomes
\be
\widehat{\balpha}=\arg\min\limits_{\balpha} g_q(\balpha,\hbR).
\label{eqn: slim_objective_modify}
\ee

The last equation can be solved by setting to zero the first complex derivative of $g_q(\balpha,\hbR)$ 
with respect to $\balpha$ to obtain
\be
-\bV^\dag\hbR^{-1}\bz+\bV^\dag\hbR^{-1}\bV\balpha + \bP^{-1}\balpha=0,
\label{eqn:derivative_alpha}
\ee
where $\bP = \diag\left(\left[ |\alpha_{1}|^{2-q},\ldots,|\alpha_{M}|^{2-q} \right] \right)$. Note that 
computing closed-form solution of the above equation is a difficult task because $\bP$ is a nonlinear function
of $\balpha$ and, hence, we resort to an iterative method. Specifically, suppose that the estimate 
of $\balpha$ at the $i$th iteration, $\tilde{\balpha}^{(i)}$ say, is available, then $\bP$ 
in \eqref{eqn:derivative_alpha} can be computed as
\be
\bP=\tilde{\bP}^{(i)} = \diag\left(\left[ |\tilde{\balpha}^{(i)}_{1}|^{2-q},\ldots,|\tilde{\balpha}^{(i)}_{M}|^{2-q} \right] \right).
\ee
Now, the estimate of $\balpha$ at the $(i+1)$th iteration is obtained as follows
\be
\tilde{\balpha}^{(i+1)} = \left[\bV^\dag\hbR^{-1}\bV + \left(\tilde{\bP}^{(i)}\right)^{-1}\right]^{-1}\bV^\dag\hbR^{-1}\bz.
\label{eqn:iteration}
\ee

As for the starting point of the above iterative procedure, we exploit the unconstrained ML estimate 
of the $l$th entry of $\balpha$, namely
\be
\tilde{\alpha}_{l}^{(0)} = \frac{\bv^\dag(\theta_l)\hbR^{-1}\bz}{\bv^\dag(\theta_l)\hbR^{-1}\bv(\theta_l)}, \; l=1,\ldots,M.
\label{eqn:initialization}
\ee
Simulation results, not reported here for brevity, have highlighted that for this specific problem, the SLIM 
algorithm almost shows no improvement after $15$ iterations. For this reason, otherwise stated, all the next numerical examples
are obtained using this number of iterations. 

So far we have neglected the impact of $q$ on $\tilde{\balpha}$, which clearly depends on the former and, hence,
we denote this estimate by $\tilde{\balpha}_q$.
The number of non-zero entries of $\tilde{\balpha}_q$ is generally larger than the actual number of targets, especially for large $q$. To further improve the sparsity of $\tilde{\balpha}_q$, the model-order selection Bayesian information criterion (BIC) can be incorporated into the procedure \cite{stoica2004mos,Stoica2004GLRTmos}. 
Precisely, the BIC rule selects the order which minimizes the following objective function
\begin{align}
\mbox{BIC}_q(h) &= -2\ln f(\bz|\tilde{\balpha}_q(h);\hbR)+3h \log(2N) \nonumber
\\
&\approx 2\left\|\hbR^{-1/2}\left(\bz-\bV\tilde{\balpha}_q(h)\right)\right\|^2 + 3h \log(2N),
\label{eqn:BIC}
\end{align}
where $h$ is an integer denoting the number of selected non-zero entries in $\tilde{\balpha}_q$ (the model order), 
$\tilde{\balpha}_q(h)$ is obtained from $\tilde{\balpha}_q$ setting to zero all the entries of $\tilde{\balpha}_q$ 
except the largest $h$ values, and the coefficient 3 represents the number of unknown real-valued target 
parameters (complex amplitude and angle). Parameter $h$ is assumed to belong to
the finite set $\{ 1,\ldots, h_{\textrm{max}}\}$, where $h_{\textrm{max}}\leq M$ is the maximum 
number of targets that are supposed to be present in the operating scenario.
The specific $h$ leading to the lowest BIC value is selected as 
an estimate of the actual number of targets for a given $q$. As a consequence, for each $q$, 
the amplitude vector estimate incorporating the BIC algorithm and the corresponding BIC objective value are given by
\begin{equation}
\begin{gathered}
\widehat{\balpha}_q=\tilde{\balpha}_q(\widehat{h}),\\
\mbox{BIC}_q=\mbox{BIC}_q(\widehat{h}),
\end{gathered}
\label{eqn:MAP+BIC estimate}
\end{equation}
respectively, where $\widehat{h}=\arg\min\limits_{h}\mbox{BIC}_q(h)$.
Summarizing, for a given $q$, the pseudocode of the BIC based SLIM (BSLIM) procedure is reported in Algorithm \ref{algorithm1}.
\begin{algorithm}[H]
	\caption{: Pseudocode of the BSLIM procedure}
	\label{algorithm1}
	\textbf{Input:} Primary datum (cell under test) $\bz$, SCM $\widehat{\bR}$, and dictionary matrix $\bV$;\\
	\textbf{Output:} Amplitude vector estimate $\widehat{\balpha}_q$, and corresponding BIC objective value $\mbox{BIC}_q$;

	\begin{algorithmic}[1]
		\STATE Initialize the amplitude vector using \eqref{eqn:initialization}, denoted as $\tilde{\balpha}_q^{(0)}$;
		\STATE Implement the iterative procedure according to \eqref{eqn:iteration} for $N_{iteration}$ times and denote the output as $\tilde{\balpha}_q$;
		\STATE Compute BIC values for different $h\in\{1,\ldots,h_{\textrm{max}}\}$ 
		using \eqref{eqn:BIC} and select the one leading to the lowest BIC value, denoted as $\widehat{h}$;
		\STATE Update the amplitude vector estimate $\widehat{\balpha}_q$ and the corresponding value $\mbox{BIC}_q$ resorting to \eqref{eqn:MAP+BIC estimate}.
	\end{algorithmic}
\end{algorithm}
As to the parameter $q$, we follow the lead of \cite{stoicaSparse} by sampling the interval $(0, 1]$ to
come up with a cardinality-$Q$ set of admissible values for $q$, which is denoted as $\Omega$. 
Then, for each $q\in \Omega$, the BSLIM procedure is carried out to obtain $Q$ BIC values indicated as $\mbox{BIC}_q,\;q\in \Omega$. The optimal $q$ value is then obtained as
\be
\widehat{q} = \arg\min_{q\in \Omega} \mbox{BIC}_q.
\ee
Finally, the user parameter free amplitude vector estimate is given by
\be
\widehat{\balpha} = \widehat{\balpha}_{\widehat{q}}.
\ee
The pseudocode of the user parameter free BSLIM algorithm is summarized in Algorithm \ref{algorithm2}.
\begin{algorithm}[H]
	\caption{: Pseudocode of the user parameter free BSLIM algorithm}
	\label{algorithm2}
	\textbf{Input:} Primary datum (cell under test) $\bz$, SCM $\widehat{\bR}$, and dictionary matrix $\bV$;\\
	\textbf{Output:} Final sparse amplitude estimate $\widehat{\balpha}$;
	
	\begin{algorithmic}[1]
		\STATE Sample the interval $(0, 1]$ to come up with a cardinality-$Q$ set of admissible values for $q$, denoted as $\Omega=\{q_1,\ldots,q_Q\}$;
		\STATE For each element of $\Omega$, implement the BSLIM procedure to get the amplitude vector estimate $\widehat{\balpha}_q$ and the corresponding BIC objective value $\mbox{BIC}_q$;
		\STATE Select the specific $q$ leading to the lowest BIC objective value and denote it as $\widehat{q}$;
		\STATE Choose $\widehat{\balpha}_{\widehat{q}}$ as the final sparse amplitude estimate and denote it as $\widehat{\balpha}$.
	\end{algorithmic}
\end{algorithm}

\section{Decision Schemes Using Sparse Amplitude Estimation}
\label{Sec:decision scheme}
This section addresses the design of decision strategies exploiting the sparse amplitude estimate. 
In this respect, a simple and intuitive way to perform the detection is to check the presence of 
a non-zero amplitude estimate indexed by $m$ which, in turn, corresponds to the nominal pointing direction.
To be more definite, denoting by $\widehat{\alpha}_m$ the entry of $\balpha$ related to the nominal
steering angle, if $|\widehat{\alpha}_m|>0$, then $H_1$ is declared. In the following, this decision strategy is referred to as SAD. 
However, due to the effect of estimation errors, the sparse amplitude estimate might be inaccurate and some of the non-zero elements might not represent the location of true targets. Besides, the SAD provides no control on the false alarm rate, which is an issue of primary concern in radar. 
In order to solve this problem, two classes of architectures are introduced in the following.

\subsection{Decision Architectures Based on the Two-Stage Paradigm}
The first class of architectures relies on the two-stage paradigm \cite{danilobook, demaiobook}. Accordingly, 
the proposed detector is formed cascading two decision schemes and the final decision is taken 
by means of a logic AND between the two stages. In this context, we combine the SAD 
with classic CFAR detectors such as Kelly's GLRT and the AMF to obtain
\be
\text{SAD-GLRT}: \left\{
\begin{aligned}
	&H_0: |\widehat{\alpha}_m|=0 \quad \text{or} \quad \Lambda_{\mbox{\tiny GLRT},m} < \eta_{\mbox{\tiny GLRT}} 
	\\
	&H_1:
	|\widehat{\alpha}_m|>0 \quad \text{and} \quad \Lambda_{\mbox{\tiny GLRT},m} > \eta_{\mbox{\tiny GLRT}} 
\end{aligned}
,
\right.
\label{eqn:SAD-GLRT}
\ee
\be
\text{SAD-AMF}: \left\{
\begin{aligned}
	&H_0: |\widehat{\alpha}_m|=0 \quad \text{or} \quad \Lambda_{\mbox{\tiny AMF},m} < \eta_{\mbox{\tiny AMF}} 
	\\
	&H_1:
	|\widehat{\alpha}_m|>0 \quad \text{and} \quad \Lambda_{\mbox{\tiny AMF},m} > \eta_{\mbox{\tiny AMF}} 
\end{aligned}
,
\right.
\label{eqn:SAD-AMF}
\ee
where we recall that $m$ is the integer indexing the nominal pointing direction, $\Lambda_{\mbox{\tiny GLRT},m}$ and $\Lambda_{\mbox{\tiny AMF},m}$ are Kelly's GLRT and AMF decision statistics, respectively, evaluated using 
the nominal steering vector $\bv(\theta_m)$, whereas $\eta_{\mbox{\tiny GLRT}}$ and $\eta_{\mbox{\tiny AMF}}$
are the detection thresholds for Kelly's GLRT and the AMF, respectively.

It is important to underline here that the actual $P_{fa}$ of SAD-GLRT and SAD-AMF is given by
\be
\begin{split}
	P_{fa,\mbox{\tiny SAD-GLRT}}&=P(|\widehat{\alpha}_m|>0,\Lambda_{\mbox{\tiny GLRT},m}>\eta_{\mbox{\tiny GLRT}}|H_0)\\&\leq P(\Lambda_{\mbox{\tiny GLRT},m}>\eta_{\mbox{\tiny GLRT}}|H_0),
\end{split}
\ee
\be
\begin{split}
	P_{fa,\mbox{\tiny SAD-AMF}}&=P(|\widehat{\alpha}_m|>0,\Lambda_{\mbox{\tiny AMF},m}>\eta_{\mbox{\tiny AMF}}|H_0)\\
	&\leq P(\Lambda_{\mbox{\tiny AMF},m}>\eta_{\mbox{\tiny AMF}}|H_0),
\end{split}
\ee
and, hence, the SAD-GLRT and SAD-AMF are bounded CFAR since Kelly's GLRT and AMF are CFAR detectors. 

\subsection{Decision Architectures Based on the Likelihood Ratio Test}
Another approach leading to detectors capable of controlling the false alarm rate consists in 
exploiting ad hoc modifications of the GLRT, where only some parameters are assumed unknown
and estimated through the maximum likelihood approach, whereas the other parameters
are replaced by suitable estimates.
In this case, we can exploit the sparse amplitude estimates returned by the previously
described estimation procedure.
To be more definite, let us denote by $m$ the integer indexing the nominal steering direction, then the likelihood ratio test (LRT) is given by
\be
\Lambda_{\mbox{\tiny LRT},m}=\frac{f(\bz,\bz_1,\cdots,\bz_K;\alpha_m,\bR,H_1)}{f(\bz,\bz_1,\cdots,\bz_K;\bR,H_0)}  
\test \eta, 
\label{eqn:LRT}
\ee
where 
\be
f(\bz,\bz_1,\cdots,\bz_K;\alpha_m,\bR,H_1)
=\left\{\frac{\exp\left[-\tr(\bR^{-1}\bT_1)\right]}{\pi^N\det(\bR)}\right\}^{K+1}
\label{eqn:H1}
\ee
and 
\be
f(\bz,\bz_1,\cdots,\bz_K;\bR,H_0)=\left\{\frac{\exp\left[-\tr(\bR^{-1}\bT_0)\right]}{\pi^N\det(\bR)}\right\}^{K+1}
\label{eqn:H0}
\ee
with
\be
\bT_1=\frac{1}{K+1}\left[(\bz-\alpha_m\bv(\theta_m))(\bz-\alpha_m\bv(\theta_m))^\dag+K\hbR\right]
\label{eqn:T1}
\ee
and
\be
\bT_0=\frac{1}{K+1}\left(\bz\bz^\dag+K\hbR\right)
\label{eqn:T0}
\ee
are the joint pdfs of the vectors $\bz,\bz_1,\cdots,\bz_K$ under $H_1$ and $H_0$, respectively. 
Note that by \eqref{eqn:LRT}, the LRT for the nominal angular position is dependent on two unknown quantities, namely
the target amplitude $\alpha_m$ and the interference covariance matrix $\bR$. 

\subsubsection{Case A (BSLIM-AMF)} 
If we replace $\alpha_m$ and $\bR$ in \eqref{eqn:H1}-\eqref{eqn:T0} with
the sparse estimate $\hat{\alpha}_m$ and the SCM $\hbR$, respectively, then the logarithm of \eqref{eqn:LRT} can be written as
\be
\begin{split}
	\Lambda_{\mbox{\tiny BSLIM-AMF},m}&=-\left(\bz-\hat{\alpha}_m\bv(\theta_m)\right)^\dag\hbR^{-1}\left(\bz-\hat{\alpha}_m\bv(\theta_m)\right)\\
	&+\bz^\dag\hbR^{-1}\bz
	\test \eta, 
\end{split}
\label{eqn:BSLIM-AMF}
\ee
which will be referred to in the following as BSLIM-AMF.

To gain some insight in \eqref{eqn:BSLIM-AMF}, let us denote the unconstrained ML estimate of $\alpha_m$ for the single target scenario as
\be
\hat{\alpha}_{\mbox{\tiny ML},m}=\frac{\bv^\dag(\theta_m)\hbR^{-1}\bz}{\bv^\dag(\theta_m)\hbR^{-1}\bv(\theta_m)},
\ee
the decision statistic of \eqref{eqn:BSLIM-AMF} can be recast as
\be
\begin{split}
	&\Lambda_{\mbox{\tiny BSLIM-AMF},m}\\
	&=-\left(\bz-\hat{\alpha}_m\bv(\theta_m)\right)^\dag\hbR^{-1}\left(\bz-\hat{\alpha}_m\bv(\theta_m)\right)+\bz^\dag\hbR^{-1}\bz\\
	&=-\left(\bv^\dag(\theta_m)\hbR^{-1}\bv(\theta_m)\right)\left|\hat{\alpha}_m-\frac{\bv^\dag(\theta_m)\hbR^{-1}\bz}{\bv(\theta_m)^\dag\hbR^{-1}\bv(\theta_m)}\right|^2\\
	&+\frac{|\bv^\dag(\theta_m)\hbR^{-1}\bz|^2}{\bv^\dag(\theta_m)\hbR^{-1}\bv(\theta_m)}\\
	&=\Lambda_{\mbox{\tiny AMF},m}-\left(\bv^\dag(\theta_m)\hbR^{-1}\bv(\theta_m)\right)\left|\hat{\alpha}_m-\hat{\alpha}_{\mbox{\tiny ML},m}\right|^2\\
	&=\Lambda_{\mbox{\tiny AMF},m}\left(1-\frac{|\hat{\alpha}_m-\hat{\alpha}_{\mbox{\tiny ML},m}|^2}{|\hat{\alpha}_{\mbox{\tiny ML},m}|^2}\right).
\end{split}
\label{eqn:BSLIM-AMF statistic}
\ee 
where $\Lambda_{\mbox{\tiny AMF},m}$ is the AMF decision statistic evaluated with the steering vector $\bv(\theta_m)$. The only difference between the decision statistics of BSLIM-AMF and AMF lies on the multiplier $\left(1-|\hat{\alpha}_m-\hat{\alpha}_{\mbox{\tiny ML},m}|^2/|\hat{\alpha}_{\mbox{\tiny ML},m}|^2\right)$. For matched or slightly mismatched signals and high signal-to-interference-plus-noise-ratio (SINR), both of $\hat{\alpha}_m$ and $\hat{\alpha}_{\mbox{\tiny ML},m}$ approach the true amplitude, and thus the BSLIM-AMF tends to the AMF. In contrast, for highly mismatched signals, $\hat{\alpha}_m$ normally equals to zero (namely, $\Lambda_{\mbox{\tiny BSLIM-AMF},m}=0$) whereas $\Lambda_{\mbox{\tiny AMF},m}$ might be large. This observation suggests that the BSLIM-AMF should outperform AMF in terms of rejecting unwanted signals.

\subsubsection{Case B (BSLIM-GLRT)}
Following the lead of Kelly's GLRT, it is possible to show that
\be
\begin{split}
	&\dmax_{\bR}\Lambda_{\mbox{\tiny LRT},m}=\left\{\frac{\det(\bT_0)}{\det(\bT_1)}\right\}^{K+1}\\
	&=\left\{\frac{K+\bz^\dag\hbR^{-1}\bz}{K+\left(\bz-{\alpha}_m\bv(\theta_m)\right)^\dag\hbR^{-1}\left(\bz-{\alpha}_m\bv(\theta_m)\right)}\right\}^{K+1}.
\end{split}
\label{eqn:MLRT}
\ee
Using the sparse estimate $\hat{\alpha}_m$ in place of $\alpha_m$ and taking the $(K+1)$st root, 
we obtain the following decision rule
\be
\frac{K+\bz^\dag\hbR^{-1}\bz}{K+\bz^\dag\hbR^{-1}\bz-\Lambda_{\mbox{\tiny BSLIM-AMF},m}}  \test \eta,
\label{eqn:MLRT2}
\ee
which is statistically equivalent to
\be
\begin{split}
	\Lambda_{\mbox{\tiny BSLIM-GLRT},m}&=\frac{\Lambda_{\mbox{\tiny BSLIM-AMF},m}}{K+\bz^\dag\hbR^{-1}\bz}\\
	&=\Lambda_{\mbox{\tiny GLRT},m}\left(1-\frac{|\hat{\alpha}_m-\hat{\alpha}_{\mbox{\tiny ML},m}|^2}{|\hat{\alpha}_{\mbox{\tiny ML},m}|^2}\right) \test \eta.
\end{split}
\label{eqn:BSLIM-GLRT}
\ee
The above architectures is referred to as BSLIM-GLRT. Similarly, 
the BSLIM-GLRT differs from Kelly's GLRT only by the 
multiplier $\left(1-|\hat{\alpha}_m-\hat{\alpha}_{\mbox{\tiny ML},m}|^2/|\hat{\alpha}_{\mbox{\tiny ML},m}|^2\right)$.

As a final remark, note that the left hand side of \eqref{eqn:BSLIM-AMF} and \eqref{eqn:BSLIM-GLRT} are 
upper bounded by the AMF and Kelly's GLRT decision statistics, respectively, the BSLIM-AMF and BSLIM-GLRT are also bounded CFAR. 

\bigskip

\begin{figure}[htb] \centering
	\subfigure[]{\includegraphics[width=0.9\columnwidth]{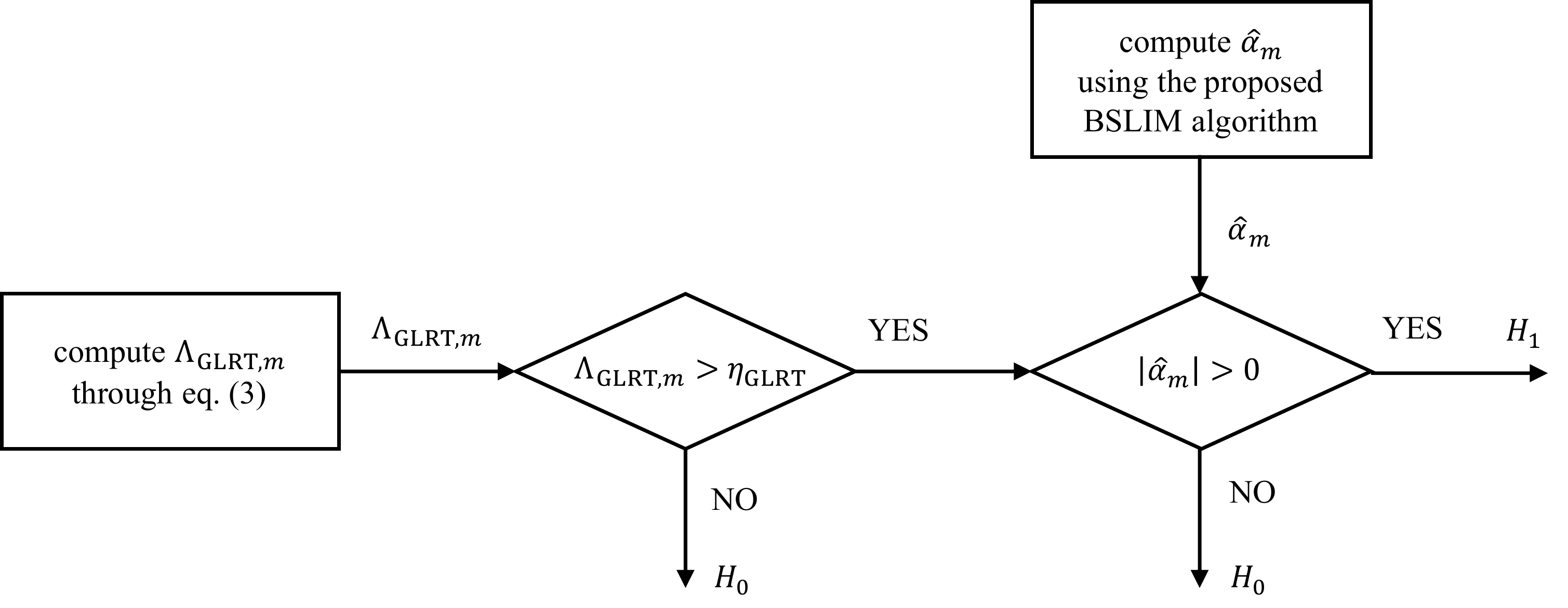}}\\
	\subfigure[]{\includegraphics[width=0.9\columnwidth]{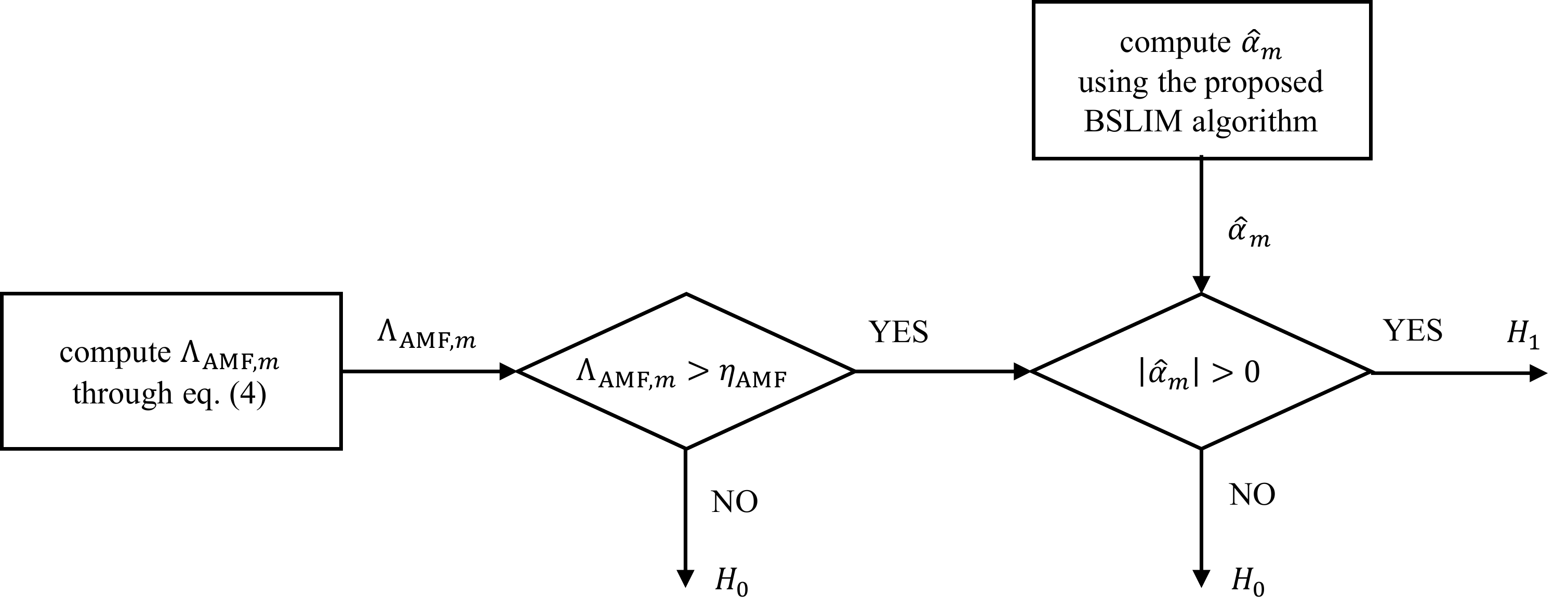}}\\
	\subfigure[]{\includegraphics[width=0.9\columnwidth]{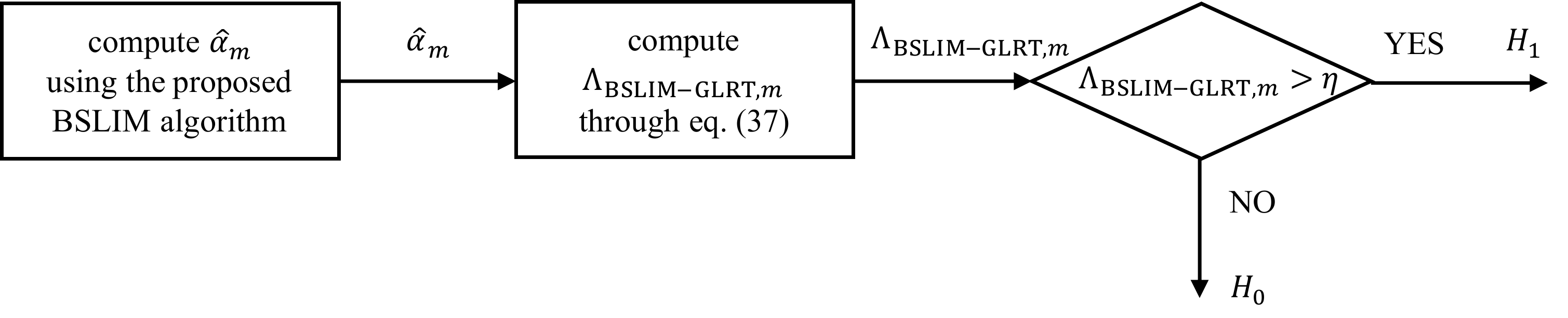}}
	\subfigure[]{\includegraphics[width=0.9\columnwidth]{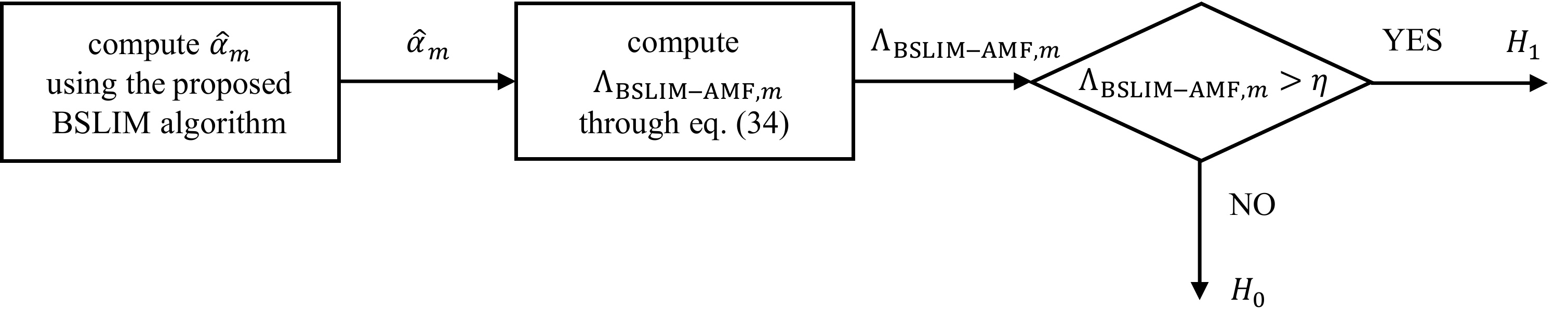}}
	\caption{Block schemes of the proposed decision architectures. 
	(a) Two-stage detector based upon Kelly's GLRT. (b) Two-stage detector based upon the AMF.
	(c)	One-stage detector based upon the GLRT. (d) One-stage detector based upon the AMF.}
	\label{block_scheme}
\end{figure}

The block-schemes of the proposed decision architectures are depicted in Fig. \ref{block_scheme}.

\section{Numerical Examples}
\label{performance assessment}
In this section, we focus on the performance assessment of the SAD-AMF, SAD-GLRT, BSLIM-AMF, 
and BSLIM-GLRT, also in comparison with well-known selective decision schemes. 
Since closed-form expressions for the $P_{fa}$ and the probability of detection ($P_d$) are not available for the new detectors, we evaluate them resorting to standard Monte Carlo techniques based on $1000/P_{fa}$ and $10^4$ independent trials, respectively. The interference is modeled as an exponentially-correlated complex Gaussian vector with one-lag correlation coefficient $\rho$, namely the $(i,j)$th element of the covariance matrix $\bR$ is given by $\rho^{|i-j|}$, $i,j=1,\cdots,N$, with $\rho=0.95$. Finally, 
the interelement spacing, namely $d$, is set to $\lambda/2$, $h\in\{1,\ldots,M\}$, and 
$q\in\Omega=\{0.01,0.1,0.2,\ldots,1\}$.

\subsection{$P_{fa}$ analysis}
\begin{figure}[htb] \centering
	\includegraphics[width=0.7\columnwidth]{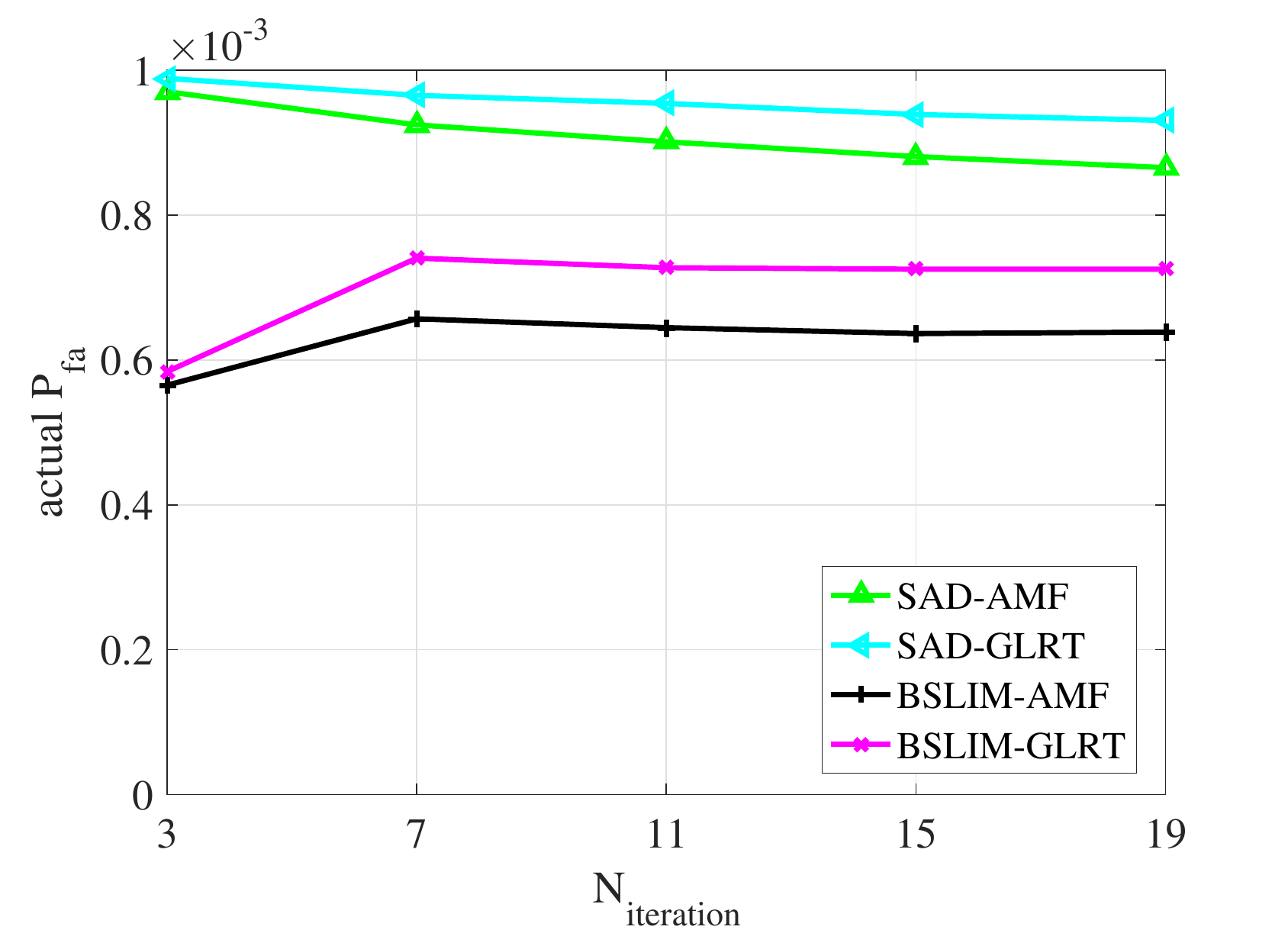}
	\caption{Actual $P_{fa}$ versus $N_{iteration}$ for $N=8$, $K=32$, and $\Delta\theta=3^{\circ}$.}
	\label{figure_pfa1}
\end{figure}

\begin{figure}[htb] \centering
	\includegraphics[width=0.7\columnwidth]{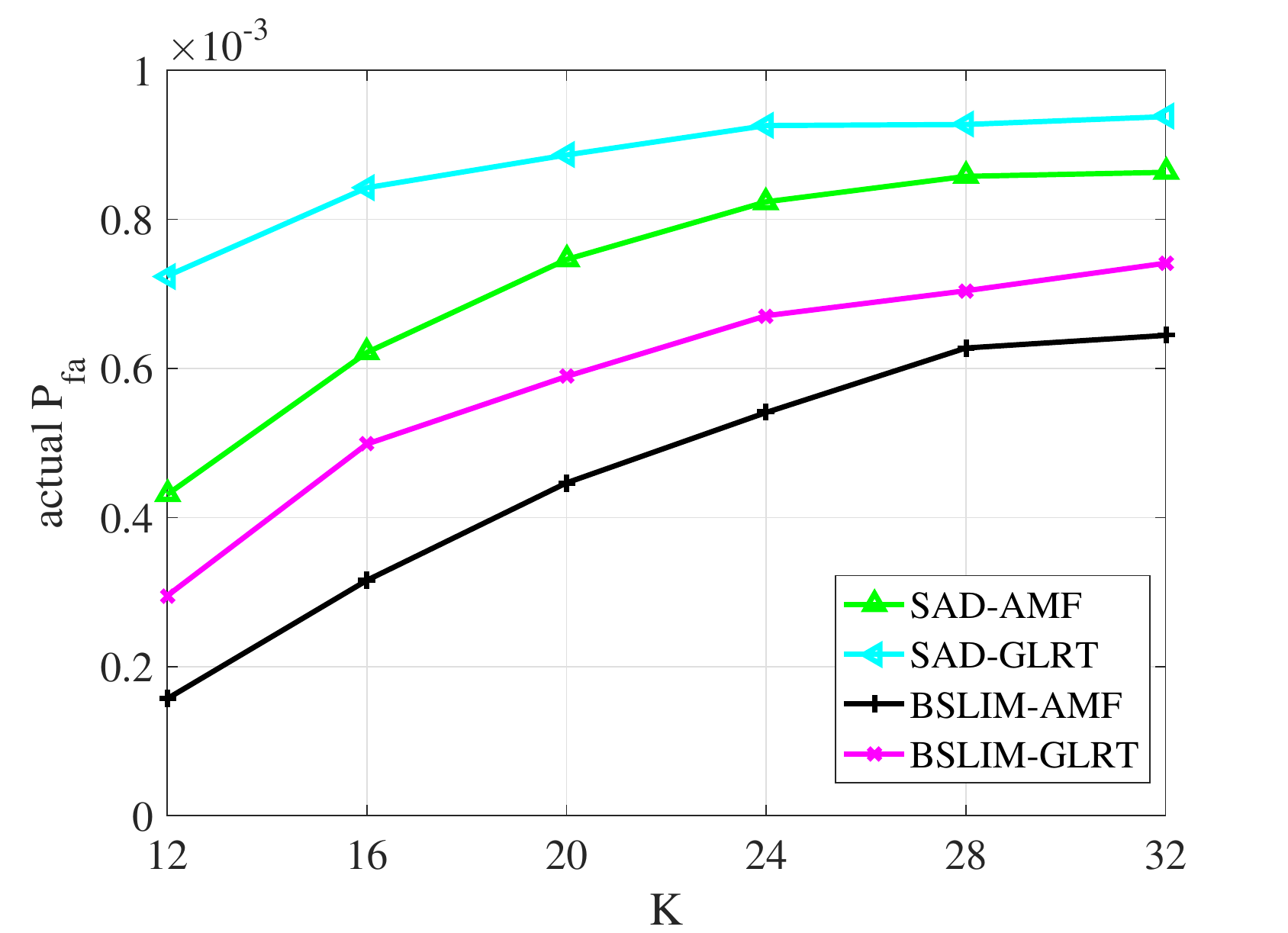}
	\caption{Actual $P_{fa}$ versus $K$ for $N=8$, $\Delta\theta=3^{\circ}$, and $N_{iteration}=15$.}
	\label{figure_pfa2}
\end{figure}

\begin{figure}[htb] \centering
	\includegraphics[width=0.7\columnwidth]{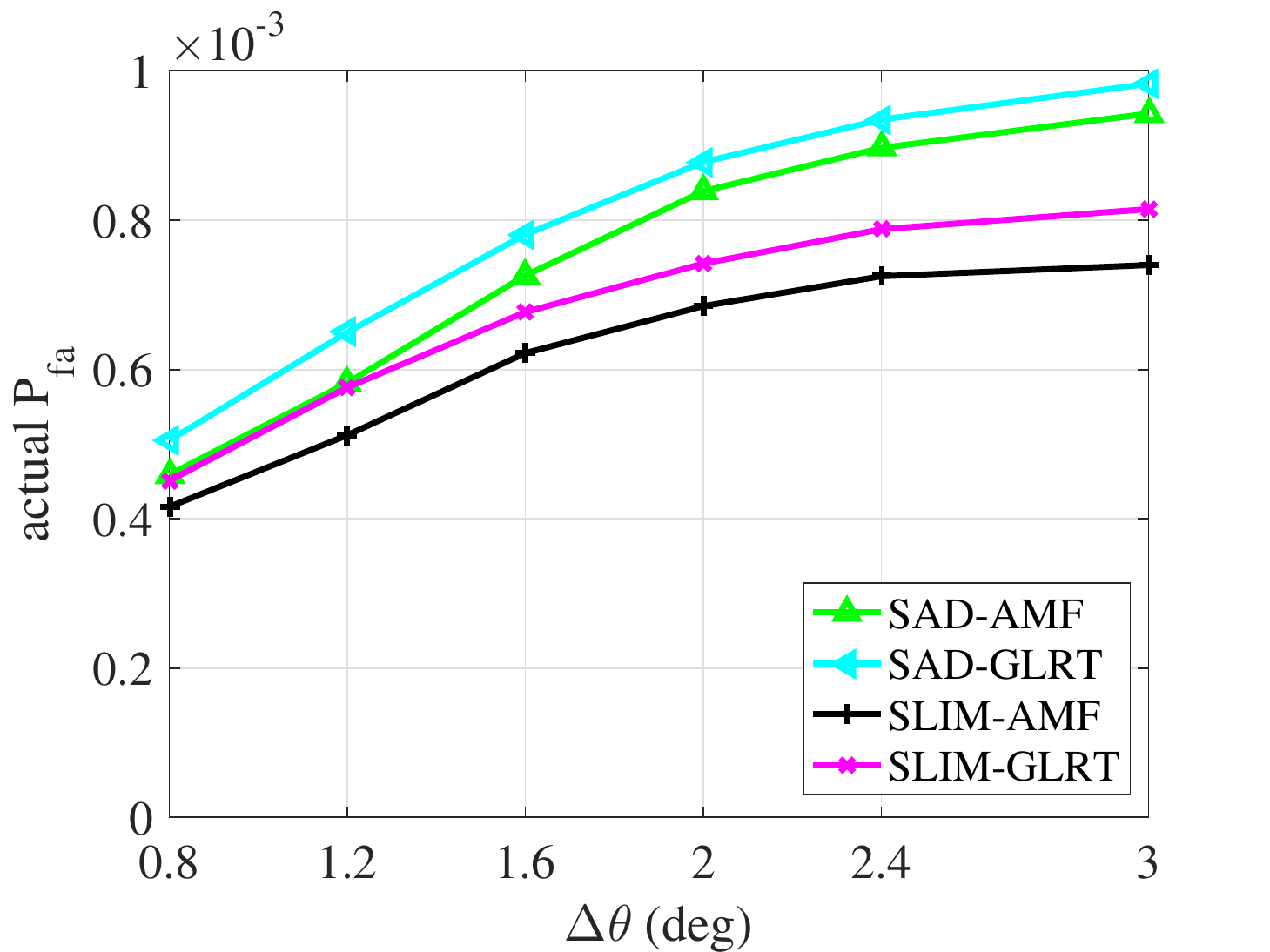}
	\caption{Actual $P_{fa}$ versus $\Delta\theta$ for $N=8$, $K=32$, and $N_{iteration}=15$.}
	\label{figure_pfa3}
\end{figure}

In Figs. \ref{figure_pfa1} to \ref{figure_pfa3}, the $P_{fa}$ behavior of the new decision schemes is studied assuming that $N=8$ and the nominal pointing direction is $0^{\circ}$. The angular region corresponding to the dictionary ranges from $-48^{\circ}$ to $48^{\circ}$ in order to include the mainbeam (whose 3 dB beamwidth is about $8.66^{\circ}$ and four sidelobes (two sidelobes on the left and two sidelobes on the right). Exploiting the bounded CFAR property, the thresholds in \eqref{eqn:SAD-GLRT}, \eqref{eqn:SAD-AMF}, \eqref{eqn:BSLIM-AMF} and \eqref{eqn:BSLIM-GLRT} are set equal to those of the AMF and Kelly's GLRT, respectively. Precisely, the thresholds in \eqref{eqn:SAD-GLRT} and \eqref{eqn:BSLIM-GLRT} are set equal to $\eta_{\mbox{\tiny GLRT}}$, which is calculated according to \eqref{eqn:Kelly_thr} whereas the thresholds in \eqref{eqn:SAD-AMF} and \eqref{eqn:BSLIM-AMF} are set to $\eta_{\mbox{\tiny AMF}}$, which is computed resorting to \eqref{eqn:AMF_thr}. Besides, $P_{fa}$ is set to $10^{-3}$ in order to limit the computational burden. Fig. \ref{figure_pfa1} shows the actual $P_{fa}$ versus the number of iterations $N_{iteration}$ used in the BSLIM procedure for $K=32$ and $\Delta\theta=3^{\circ}$ (namely, $M=33$). The curves confirm that all the new decision rules possess the bounded CFAR property. Moreover, for the chosen parameters, the $P_{fa}$ value of them almost keeps unaltered when $N_{iteration}\geq 11$. The $P_{fa}$ values versus $N_{iteration}$ for other parameters have been also evaluated. Such simulations, not reported here for brevity, highlight that the $P_{fa}$ value for the considered decision schemes almost shows no changes when $N_{iteration}\geq15$. The bounded CFAR property of the new decision architectures is further validated in Fig. \ref{figure_pfa2}, where the actual $P_{fa}$ versus $K$ is plotted for $\Delta\theta=3^{\circ}$ and $N_{iteration}=15$. The plot shows that larger $K$ generally leads to $P_{fa}$ values closer and closer to the nominal $P_{fa}$ value and the $P_{fa}$ behavior shows small changes when $K$ is sufficiently large. Furthermore, for a given $K$, the $P_{fa}$ value of the SAD-GLRT is the closest to the nominal one among the considered detectors while that of the BSLIM-AMF exhibits the greatest deviation. Fig. \ref{figure_pfa3} plots the actual $P_{fa}$ versus $\Delta\theta$ for $K=32$ and $N_{iteration}=15$. The curves 
point out that the estimated $P_{fa}$ value approaches the nominal $P_{fa}$ value as $\Delta\theta$ increases. It is worth noticing that the $P_{fa}$ deviation is generally smaller than one order of magnitude, i.e., 
$0.1\times$ nominal $P_{fa}$ value. As a consequence, in practical applications, the angular separation 
$\Delta\theta$ could be properly selected in order to ensure that the $P_{fa}$ deviation is smaller than 
the maximum tolerable value.

\subsection{$P_d$ analysis}

In the sequel, the $P_d$ behavior of the proposed decision schemes is analyzed for both matched and mismatched signals assuming $P_{fa}=10^{-3}$, the nominal pointing direction is $0^{\circ}$ and $N_{iteration}=15$, 
also in comparison with the AMF \cite{robey1992cfar}, Kelly's GLRT \cite{kelly1986adaptive}, the Rao detector (RAO) \cite{demaio2007rao}, the W-ABORT \cite{bandiera2008abort}, and the ACE \cite{conte1995ace, scharf1999ace}. For the reader ease, the expressions of the RAO, the W-ABORT, and the ACE are given by
\be
\Lambda_{\mbox{\tiny RAO}}=\frac{\left|\bv^\dag(\theta_{m})\widehat{\bS}_1^{-1}\bz\right|^2}{\bv^\dag(\theta_{m})\widehat{\bS}_1^{-1}\bv(\theta_{m})}, 
\ee
\be
\Lambda_{\mbox{\tiny WA}} =\frac{\left(K+\bz^\dag\hbR^{-1}\bz\right)^{-1}}{\left[1-\frac{\left|\bv^\dag(\theta_{m})\hbR^{-1}\bz\right|^2}{\left(\bv^\dag(\theta_{m})\hbR^{-1}\bv(\theta_{m})\right)\left(K+\bz^\dag\hbR^{-1}\bz\right)}\right]^2},
\ee
and
\be
\Lambda_{\mbox{\tiny ACE}}=\frac{\left|\bv^\dag(\theta_{m})\hbR^{-1}\bz\right|^2}{\left(\bv^\dag(\theta_{m})\hbR^{-1}\bv(\theta_{m})\right)\left(\bz^\dag\hbR^{-1}\bz\right)}, 
\ee
respectively, where $\widehat{\bS}_1=\bz\bz^\dag+K\hbR$. According to the decision rules \eqref{eqn:SAD-GLRT}, \eqref{eqn:SAD-AMF}, \eqref{eqn:BSLIM-AMF statistic} and \eqref{eqn:BSLIM-GLRT}, the detection performance is affected by the sparse amplitude estimate. The latter is dependent on the following parameters: the dictionary $\bV$ (including the number of columns $M$, the size of each column $N$ and the angular separation between adjacent columns $\Delta\theta$), the interference covariance matrix $\bR$, the number of secondary data $K$, the true target angle $\theta_t$, and the SINR, which is defined as
\be
\text{SINR}=|\alpha|^2\bv(\theta_t)^\dag\bR^{-1}\bv(\theta_t),
\ee  
where we recall that $\theta_t$ is the actual AOA of the target. Moreover, for a specific target, 
the $P_d$ value obtained using the SAD-AMF and BSLIM-AMF is smaller than or equal to that of the AMF, 
whereas the upper bound for the $P_d$ of SAD-GLRT and BSLIM-GLRT is represented by that of Kelly's GLRT.

It is now worth to underline that as claimed in \cite{Candes2008cs}, typically, to obtain a satisfactory sparse amplitude estimate exploiting sparse recovery algorithms, under the additive white Gaussian noise assumption, the dictionary is required to meet some requirements such as the restricted isometry property (RIP) or low coherence. The signal model \eqref{eqn:z} in Section \ref{Sec:Problem_Formulation} can be transformed by whitening with respect to the interference covariance matrix, namely
\be
\bz_w=\bV_w\balpha+\bn_w,
\ee
where $\bV_w=\bR^{-1/2}\bV$ and $\bn_w=\bR^{-1/2}\bn$ represent the whitened dictionary and interference component, respectively. As a consequence, the coherence of the dictionary is given by \cite{Candes2008cs}
\be
\begin{split}
	\mu(\bV_w)&=\arg\max_{1\leq i,j \leq M, i\neq j}\left|\frac{\bv_{w,i}^\dag\bv_{w,j}}{\|\bv_{w,i}\|\|\bv_{w,j}\|}\right|\\
	&=\arg\max_{1\leq i,j \leq M, i\neq j}\left|\frac{\bv_i^\dag\bR^{-1}\bv_j}{\|\bR^{-1/2}\bv_i\|\|\bR^{-1/2}\bv_j\|}\right|,
\end{split}
\ee
where $\bv_{w,i}$ and $\bv_{i}$, $i=1,\cdots,M,$ denote the $i$th column of $\bV_w$ and $\bV$, respectively.
In addition, the coherence corresponding to the $m$th azimuth bin can be defined as
\be
\mu(\bV_w,\theta_m)=\arg\max_{\substack{1\leq i \leq M\\ \bv_i\neq \bv(\theta_m)}}\left|\frac{\bv_i^\dag\bR^{-1}\bv(\theta_m)}{\|\bR^{-1/2}\bv_i\|\|\bR^{-1/2}\bv(\theta_m)\|}\right|.
\label{miu}
\ee
In order to obtain a good estimate of $\alpha_m$, $m=1,\ldots,M$, $\mu(\bV_w,\theta_m)$ should be small. 
  
\begin{figure}[htb] \centering
	\subfigure[]{\includegraphics[width=0.49\columnwidth]{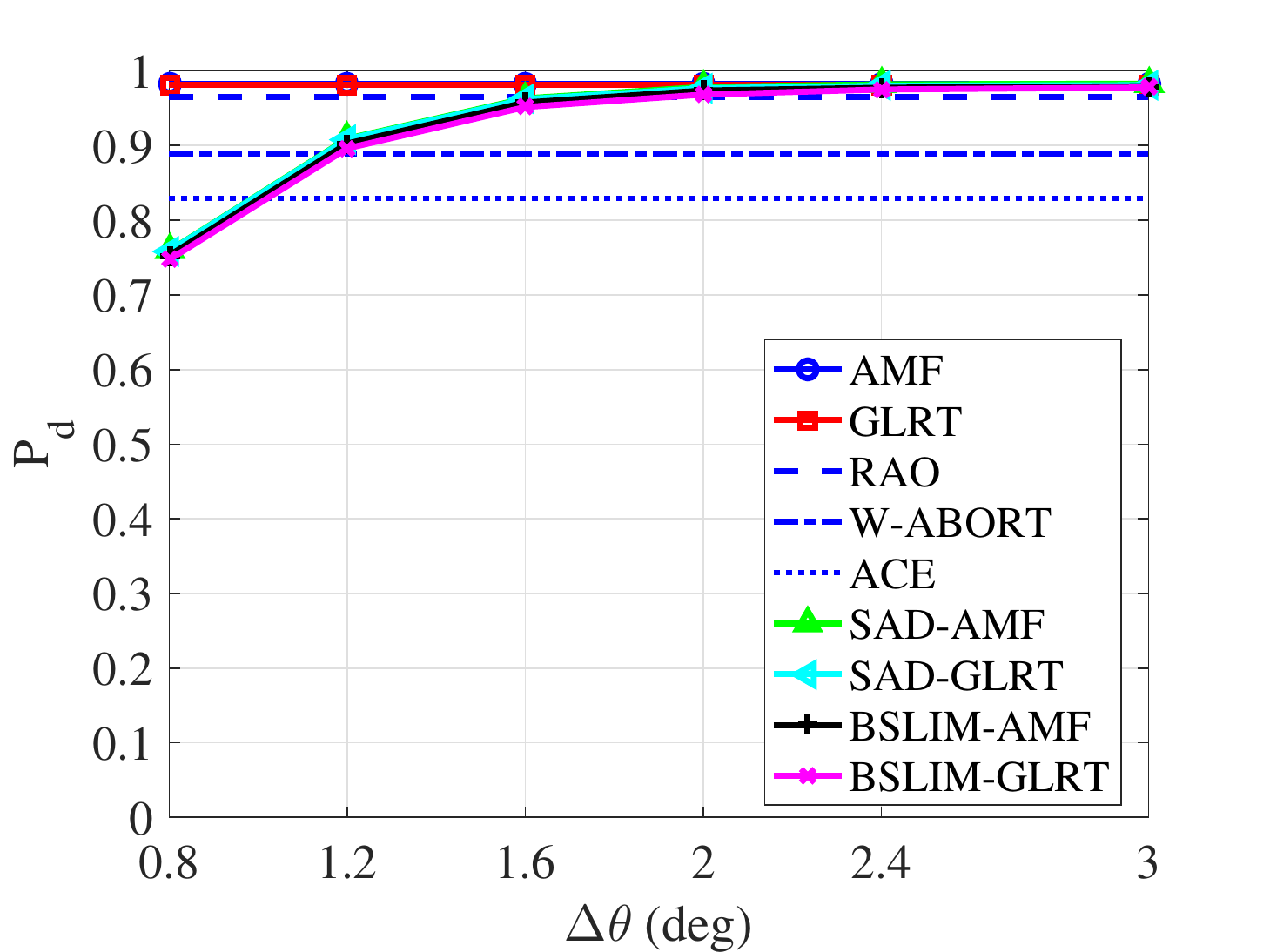}}
	\subfigure[]{\includegraphics[width=0.49\columnwidth]{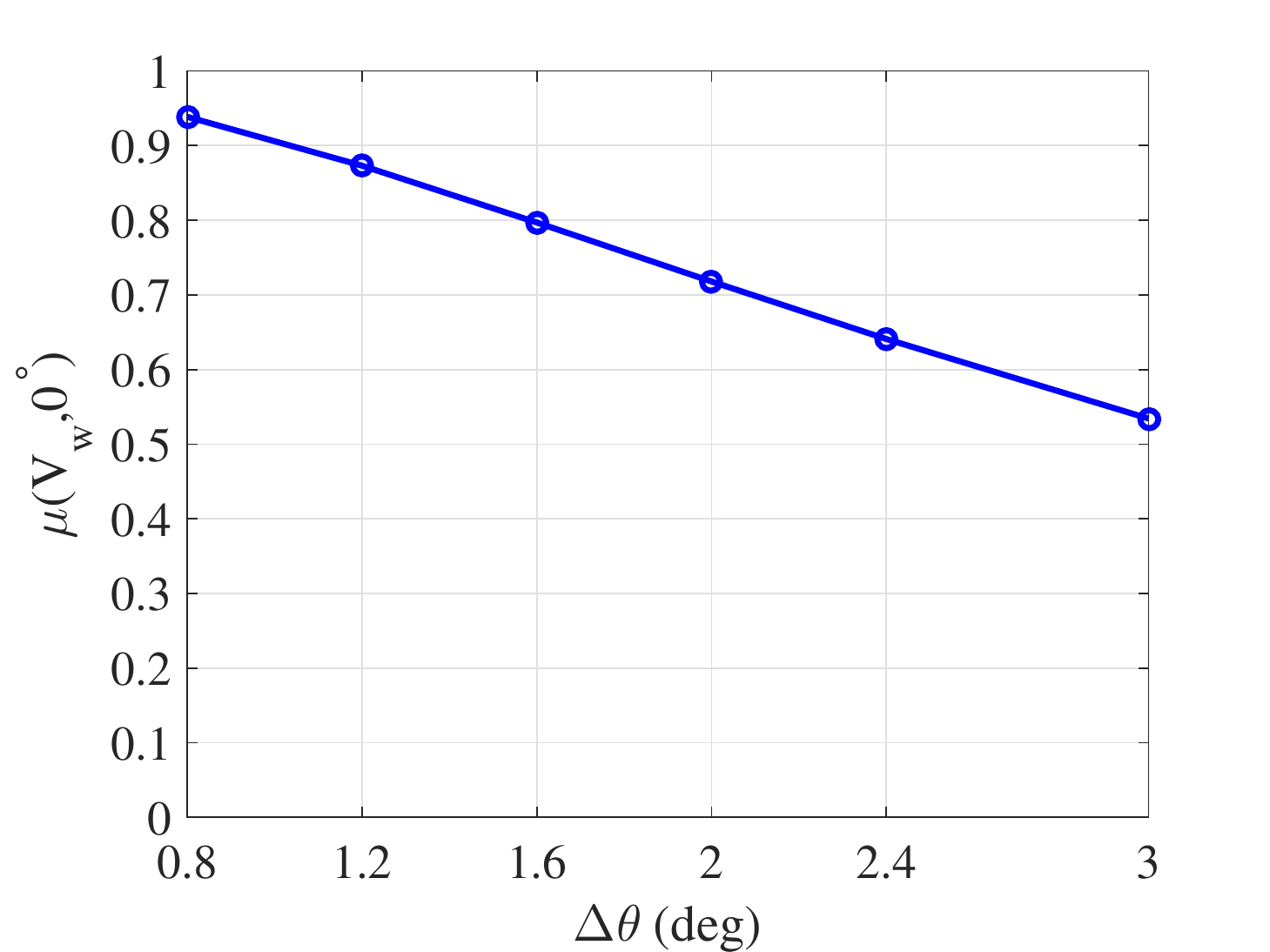}}
	\caption{$P_d$ behavior and $\mu(\bV_w,0^{\circ})$ versus $\Delta\theta$ for $N=8$, $K=32$, $\text{SINR}=14$ dB, and $\theta_t=0^{\circ}$. (a) $P_d$. (b) $\mu(\bV_w,0^{\circ})$.}
	\label{figure_Pd_sep_8}
\end{figure}
\begin{figure}[htb] \centering
	\subfigure[]{\includegraphics[width=0.49\columnwidth]{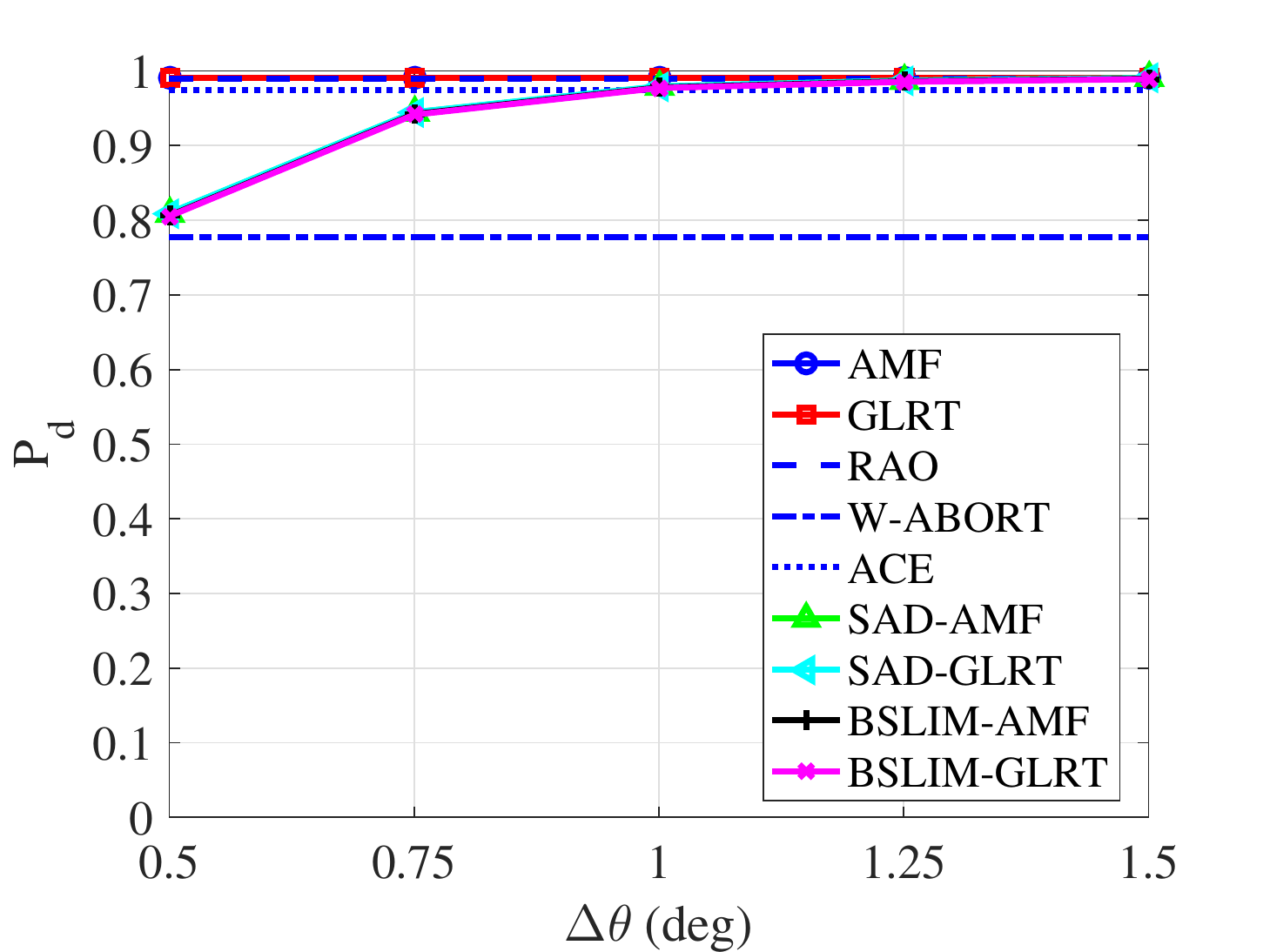}}
	\subfigure[]{\includegraphics[width=0.49\columnwidth]{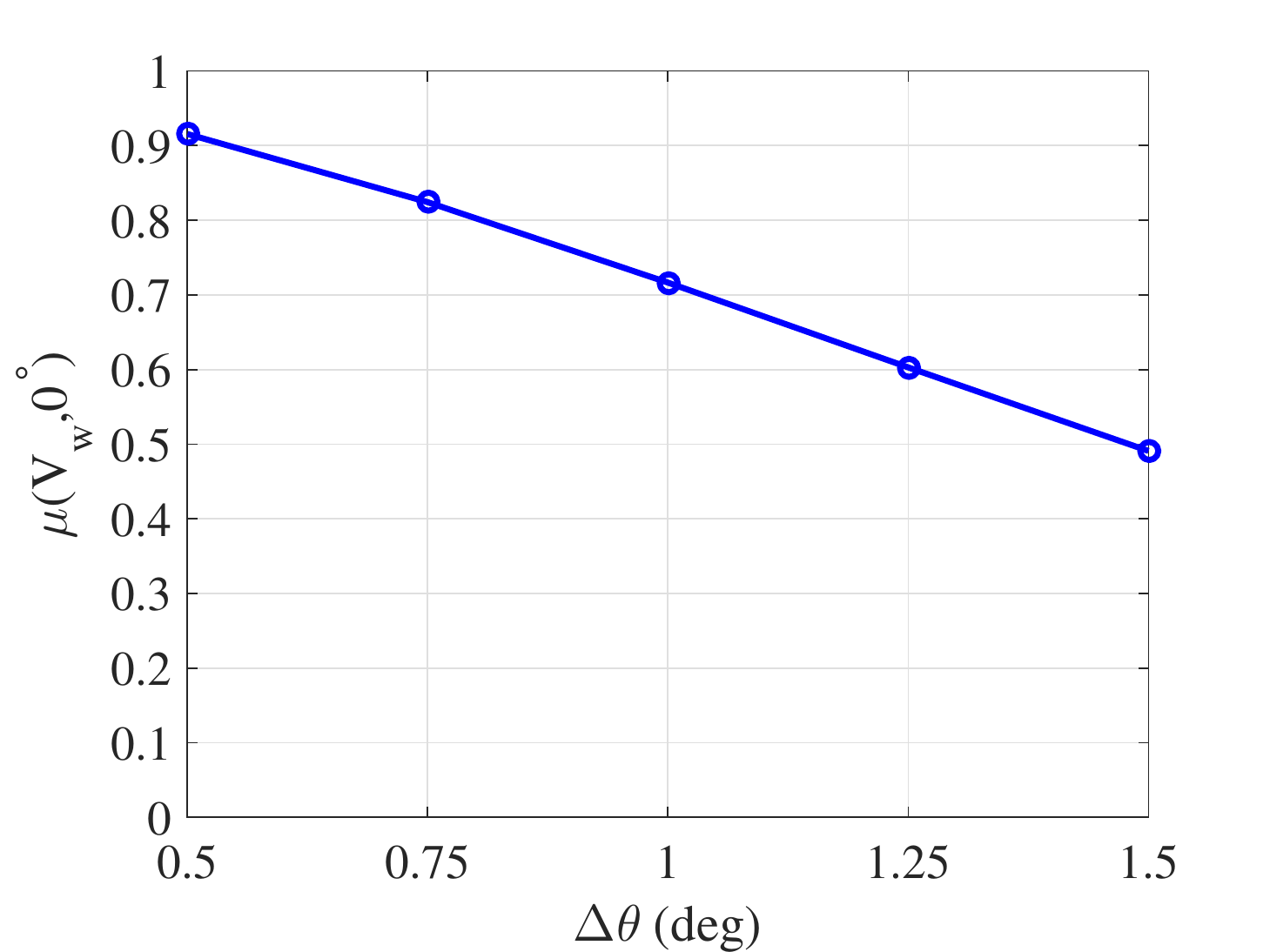}}
	\caption{$P_d$ behavior and $\mu(\bV_w,0^{\circ})$ versus $\Delta\theta$ for $N=24$, $K=96$, $\text{SINR}=14$ dB, and $\theta_t=0^{\circ}$. (a) $P_d$. (b) $\mu(\bV_w,0^{\circ})$.}
	\label{figure_Pd_sep_24}
\end{figure}
\begin{figure}[htb] \centering
	\subfigure[]{\includegraphics[width=0.49\columnwidth]{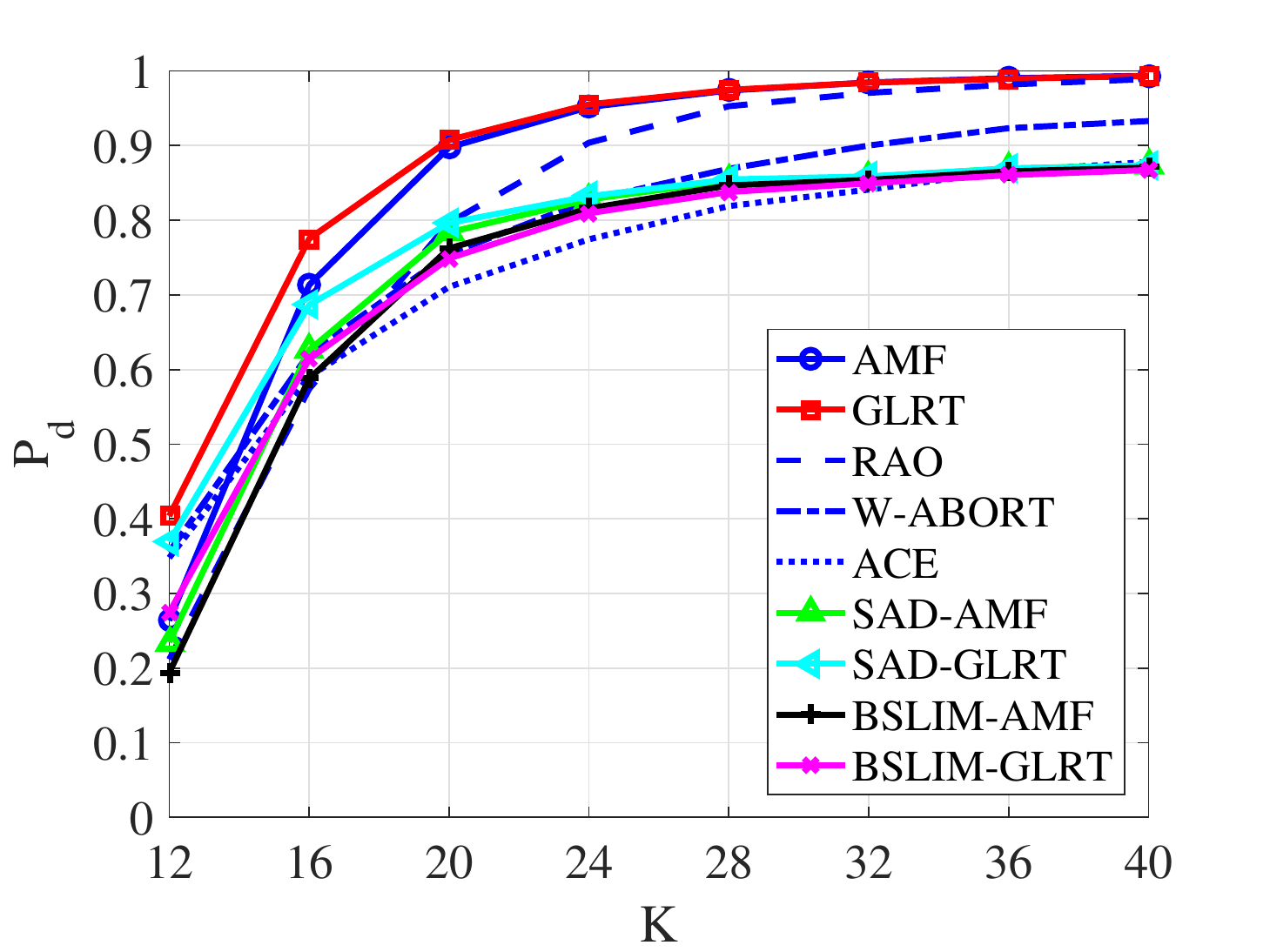}}
	\subfigure[]{\includegraphics[width=0.49\columnwidth]{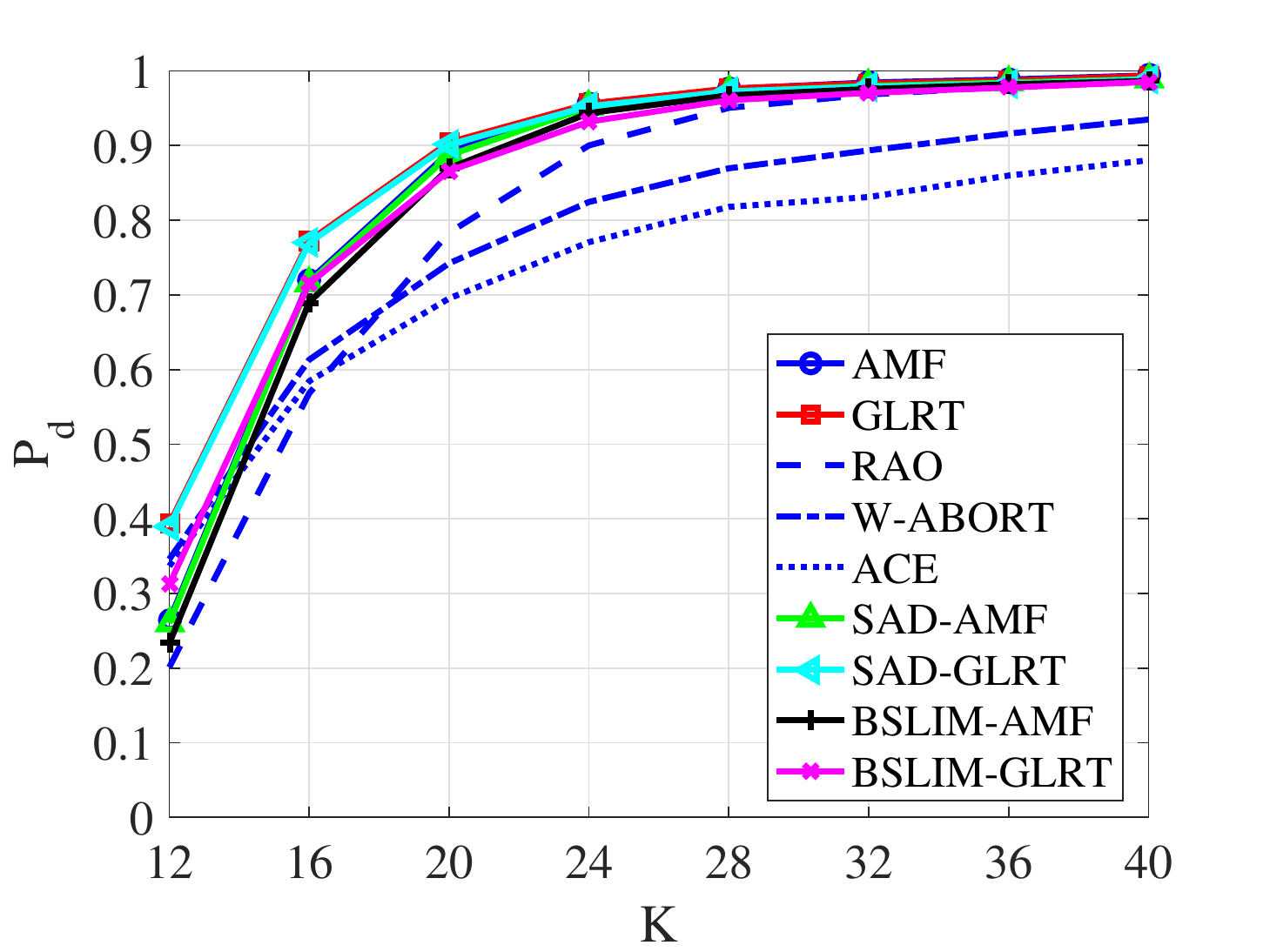}}
	\caption{$P_d$ versus $K$ for $N=8$, $\text{SINR}=14$ dB, $\theta_t=0^{\circ}$ and different $\Delta\theta$ values. (a) $\Delta\theta=1^{\circ}$. (b) $\Delta\theta=2^{\circ}$.}
	\label{figure_Pd_K}
\end{figure}
\begin{figure}[htb] \centering
	\subfigure[]{\includegraphics[width=0.49\columnwidth]{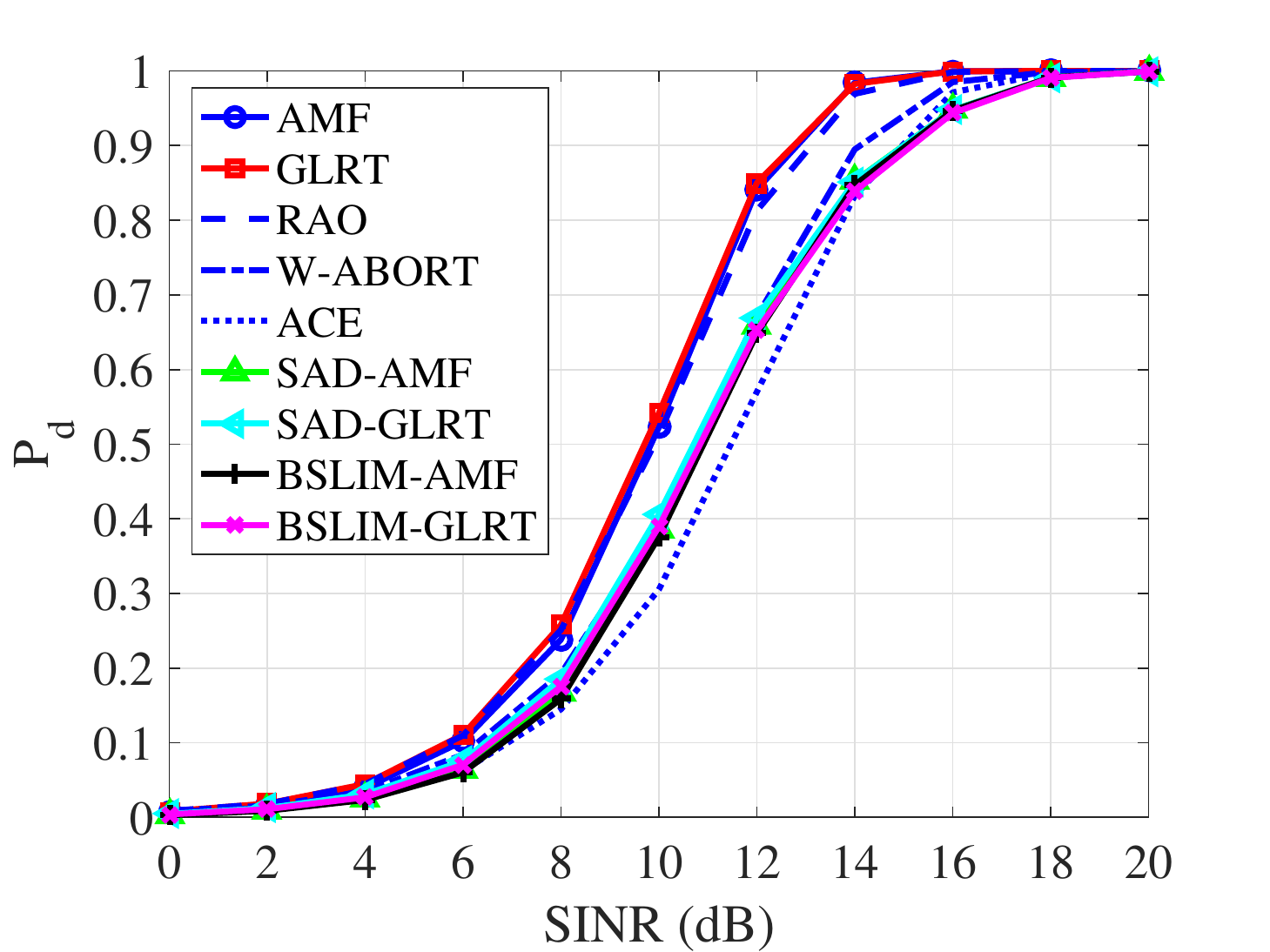}}
	\subfigure[]{\includegraphics[width=0.49\columnwidth]{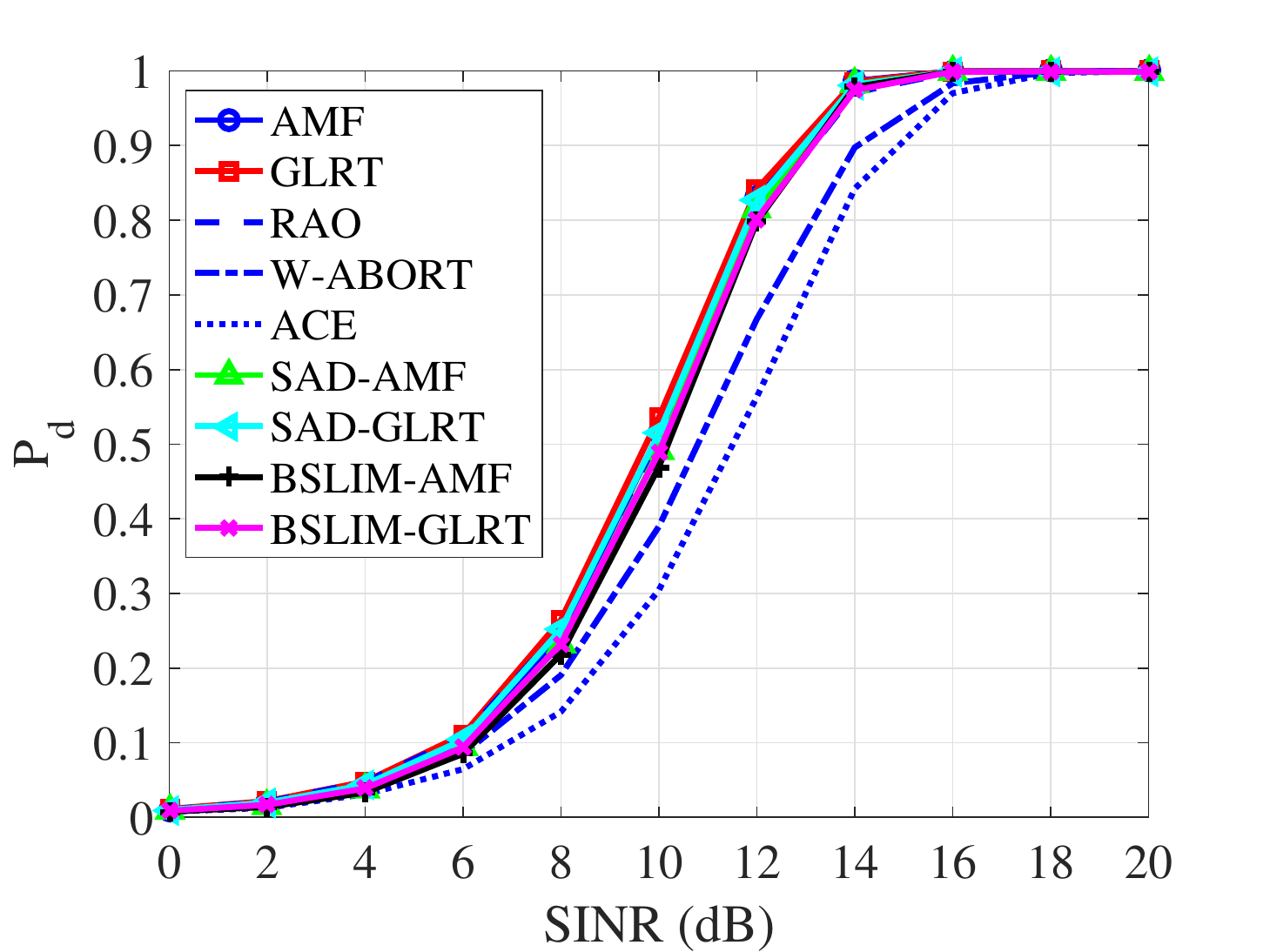}}
	\caption{$P_d$ versus SINR for $N=8$, $K=32$, $\theta_t=0^{\circ}$ and two values of $\Delta\theta$. (a) $\Delta\theta=1^{\circ}$. (b) $\Delta\theta=2^{\circ}$.}
	\label{figure_Pd_SNR}
\end{figure}

Figs. \ref{figure_Pd_sep_8} to \ref{figure_Pd_SNR} study $P_d$ under the perfect matching condition, namely, the true target angle coincides with the nominal pointing direction. In particular, Fig. \ref{figure_Pd_sep_8} plots $P_d$ versus $\Delta\theta$ for $N=8$, $K=32$, and $\text{SINR}=14$ dB. The coherence corresponding to the nominal pointing angle $\mu(\bV_w,0^{\circ})$ computed according to \eqref{miu} is also provided. The plots highlight that, for the considered parameters, the best performance is achieved by the AMF, Kelly's GLRT and RAO and their $P_d$ value is approximately 1 whereas the $P_d$ values for W-ABORT and ACE are 0.89 and 0.83, respectively. In contrast, the $P_d$ value of the new detectors increases as $\Delta\theta$ increases. This is due to the fact that the lower the coherence, the more accurate the sparse amplitude estimate and thus improved $P_d$ performance can be achieved. Moreover, the SAD-AMF, SAD-GLRT, BSLIM-AMF and BSLIM-GLRT almost share the same performance\footnote{This is not a general case. Simulations highlight that the behavior of the four considered decision schemes might exhibit a significant gap when $K$ is small (see, for instance, Fig. \ref{figure_Pd_K}).}. In Fig. \ref{figure_Pd_sep_24}, $P_d$ and coherence are plotted versus $\Delta\theta$ for $N=24$, $K=96$, and $\text{SINR}=14$ dB. In this situation, the selected angular region, again including the mainbeam (whose 3 dB beamwidth is now about $2.88^{\circ}$) and four sidelobes, ranges from $-15^{\circ}$ to $15^{\circ}$. The plots highlight that for a fixed $\Delta\theta$, a larger value of $N$ leads to a smaller value of $\mu(\bV_w,0^{\circ})$. Besides, for this specific case, $P_d$ loss of the proposed strategies (the loss of the new detectors as compared to their respective counterparts, i.e., AMF for SAD-AMF and BSLIM-AMF, Kelly's GLRT for SAD-GLRT and BSLIM-GLRT) approximately reaches 0 when $\Delta\theta\geq 1.5$ (namely, $\mu(\bV_w,0^{\circ})\leq 0.4909$). Additional experiments have been conducted for other parameter settings, whose results are not reported here for brevity, and the results confirm that when $\mu(\bV_w,0^{\circ})\leq 0.5$ and $K\geq 4N$, the SAD-AMF, SAD-GLRT, BSLIM-AMF and BSLIM-GLRT almost ensure the same performance as their respective counterparts.

In Fig. \ref{figure_Pd_K}, the effects of $K$ on $P_d$ are analyzed, to this end, $P_d$ versus $K$ is plotted for $N=8$ and $\text{SINR}=14$ dB. Two values of $\Delta\theta$ are considered: (a) $\Delta\theta=1^{\circ}$; (b) $\Delta\theta=2^{\circ}$. The plots highlight that for all the detectors, the larger the $K$, the better the $P_d$. Moreover, for $K\geq4N$, the AMF, Kelly's GLRT and RAO approximately share the same performance whereas SAD-AMF, SAD-GLRT, BSLIM-AMF and BSLIM-GLRT curves are clustered together. Finally, for the SAD-AMF, SAD-GLRT, BSLIM-AMF and BSLIM-GLRT, when the coherence is high (case (a)), the $P_d$ loss is generally non negligible. In contrast, when the coherence is sufficiently low (case (b)), the performance degradation is usually acceptable. As a consequence, to ensure a good detection performance for matched signals, the coherence should be low. This property is further confirmed in Fig. \ref{figure_Pd_SNR}, where $P_d$ versus SINR for $N=8$, $K=32$, and two values of $\Delta\theta$: (a) $\Delta\theta=1^{\circ}$; (b) $\Delta\theta=2^{\circ}$ is plotted.

\begin{figure}[htb] \centering
	\subfigure[]{\includegraphics[width=0.49\columnwidth]{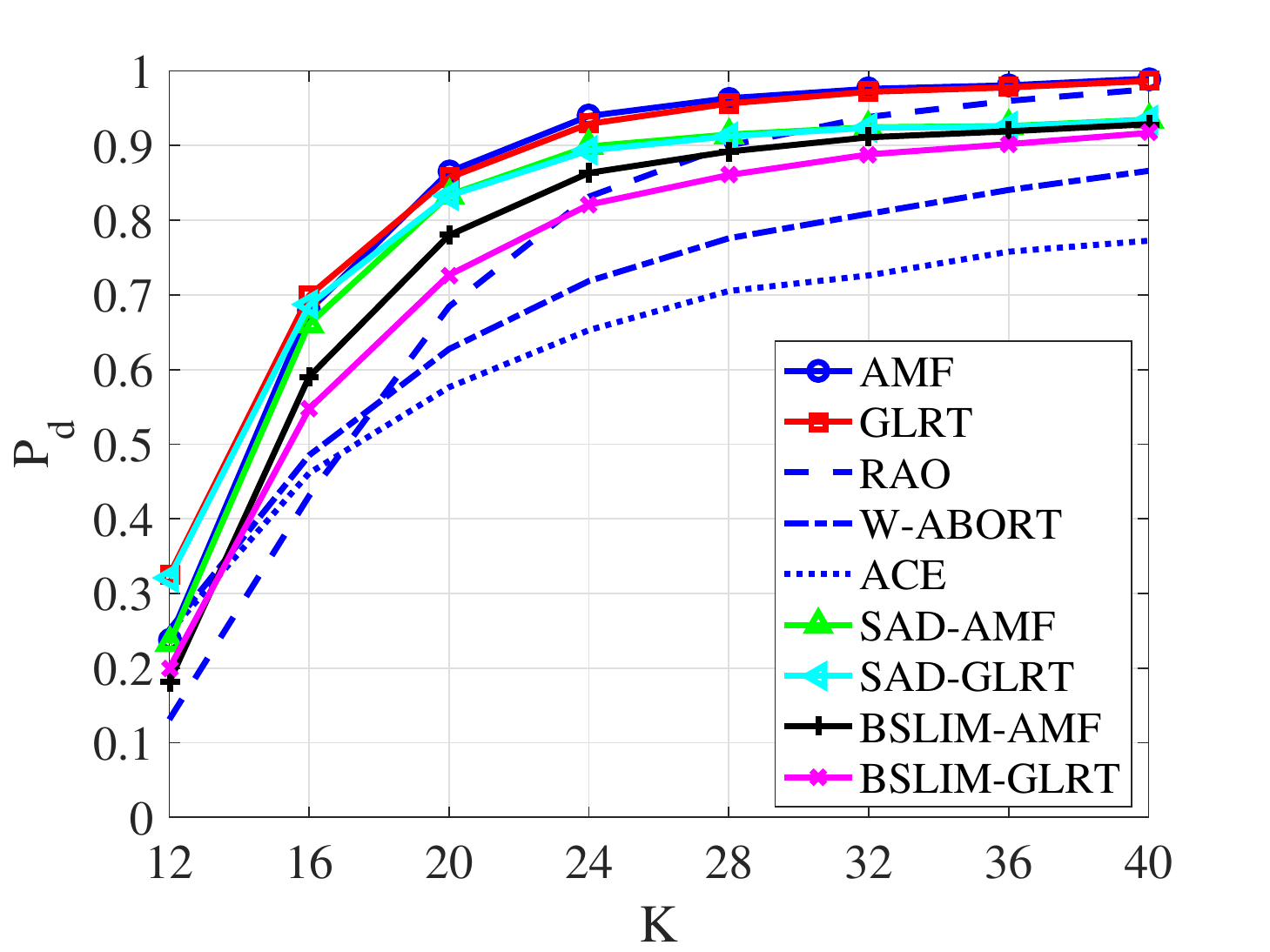}}
	\subfigure[]{\includegraphics[width=0.49\columnwidth]{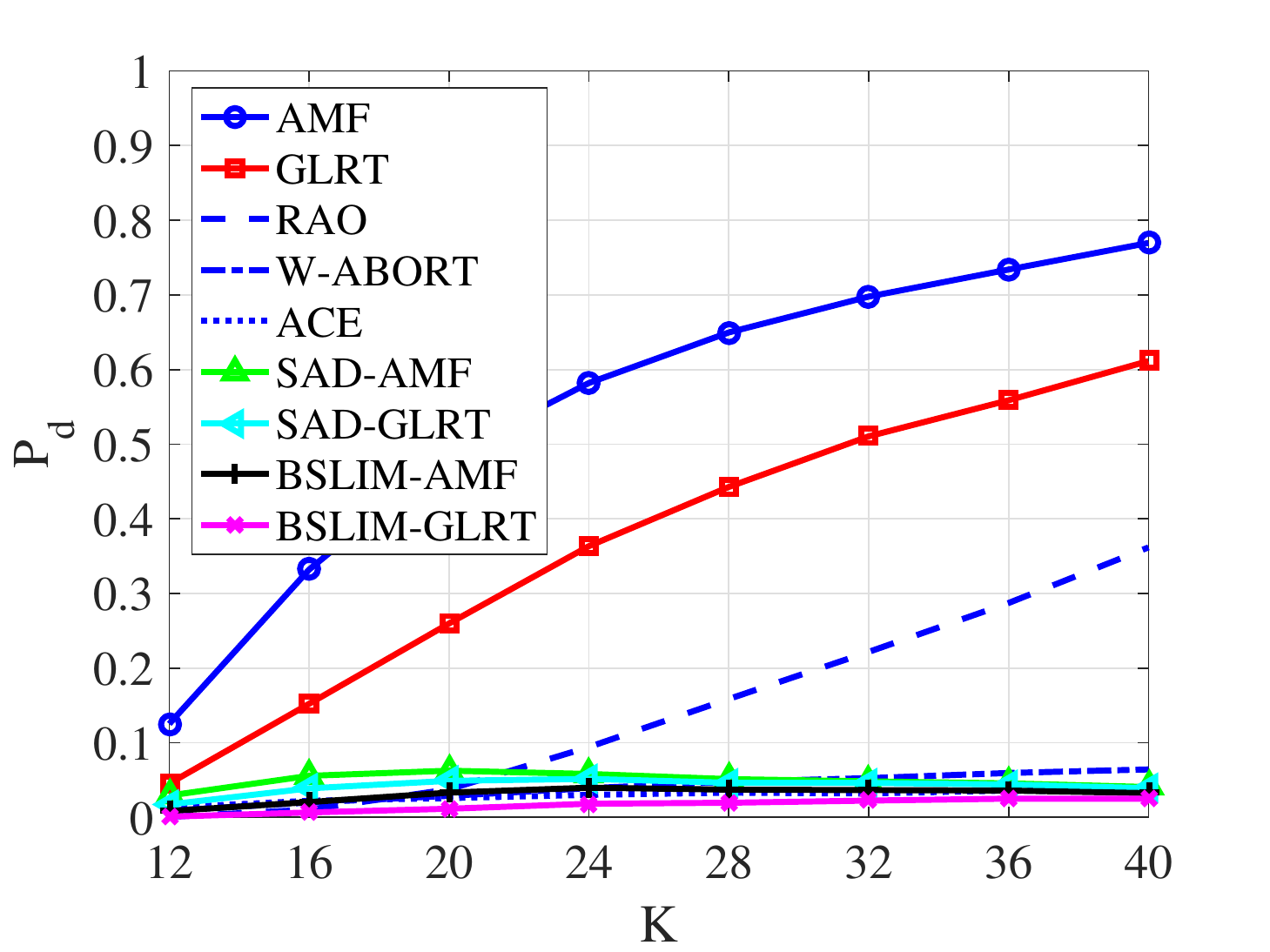}}
	\caption{$P_d$ versus $K$ for $N=8$, $\Delta\theta=2^{\circ}$, $\text{SINR}=14$ dB and two $\theta_t$ values. (a) $\theta_t=0.5^{\circ}$. (b) $\theta_t=2^{\circ}$.}
	\label{mismatched_Pd_K}
\end{figure}

\begin{figure}[htb] \centering
	\subfigure[]{\includegraphics[width=0.49\columnwidth]{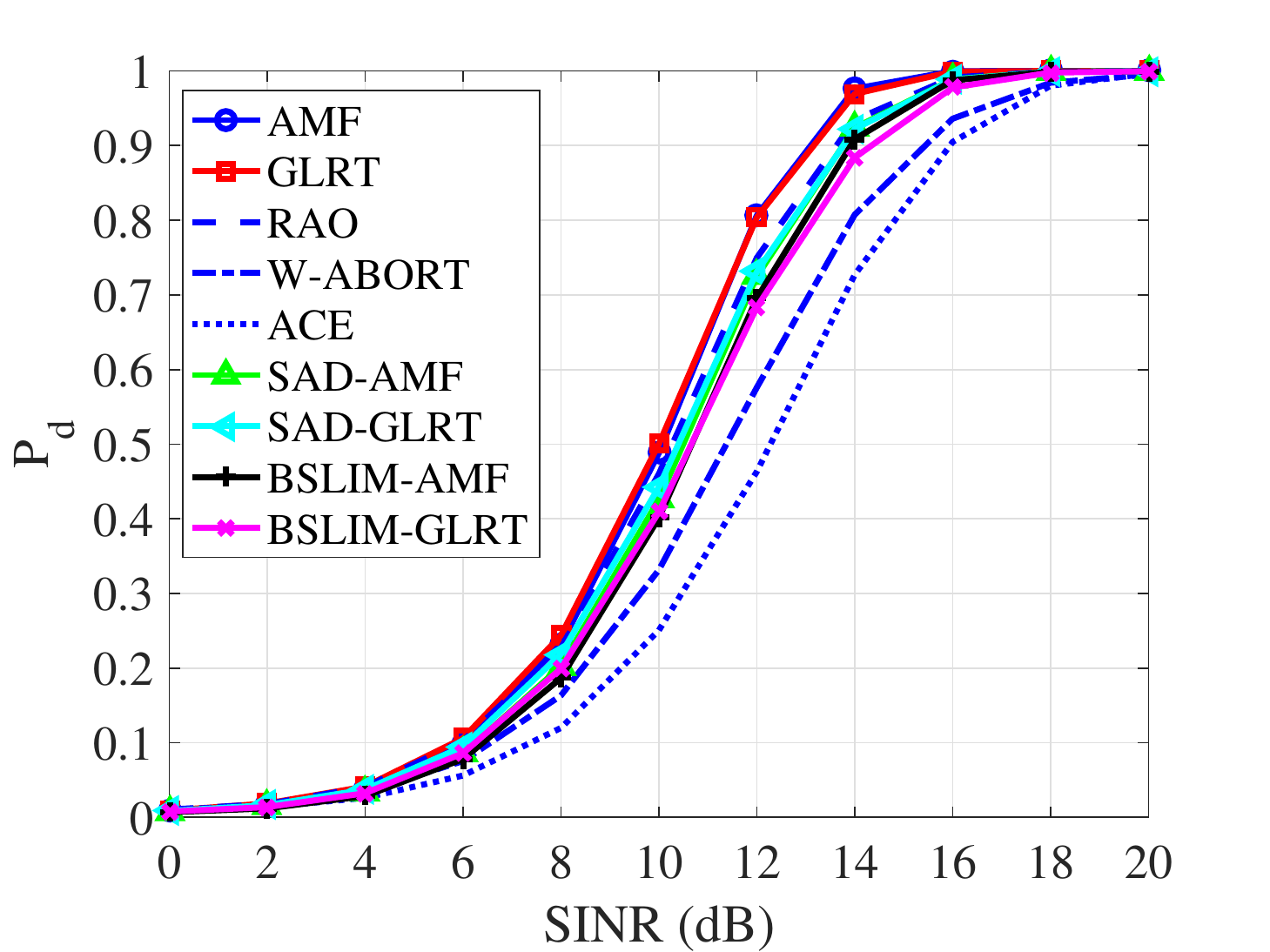}}
	\subfigure[]{\includegraphics[width=0.49\columnwidth]{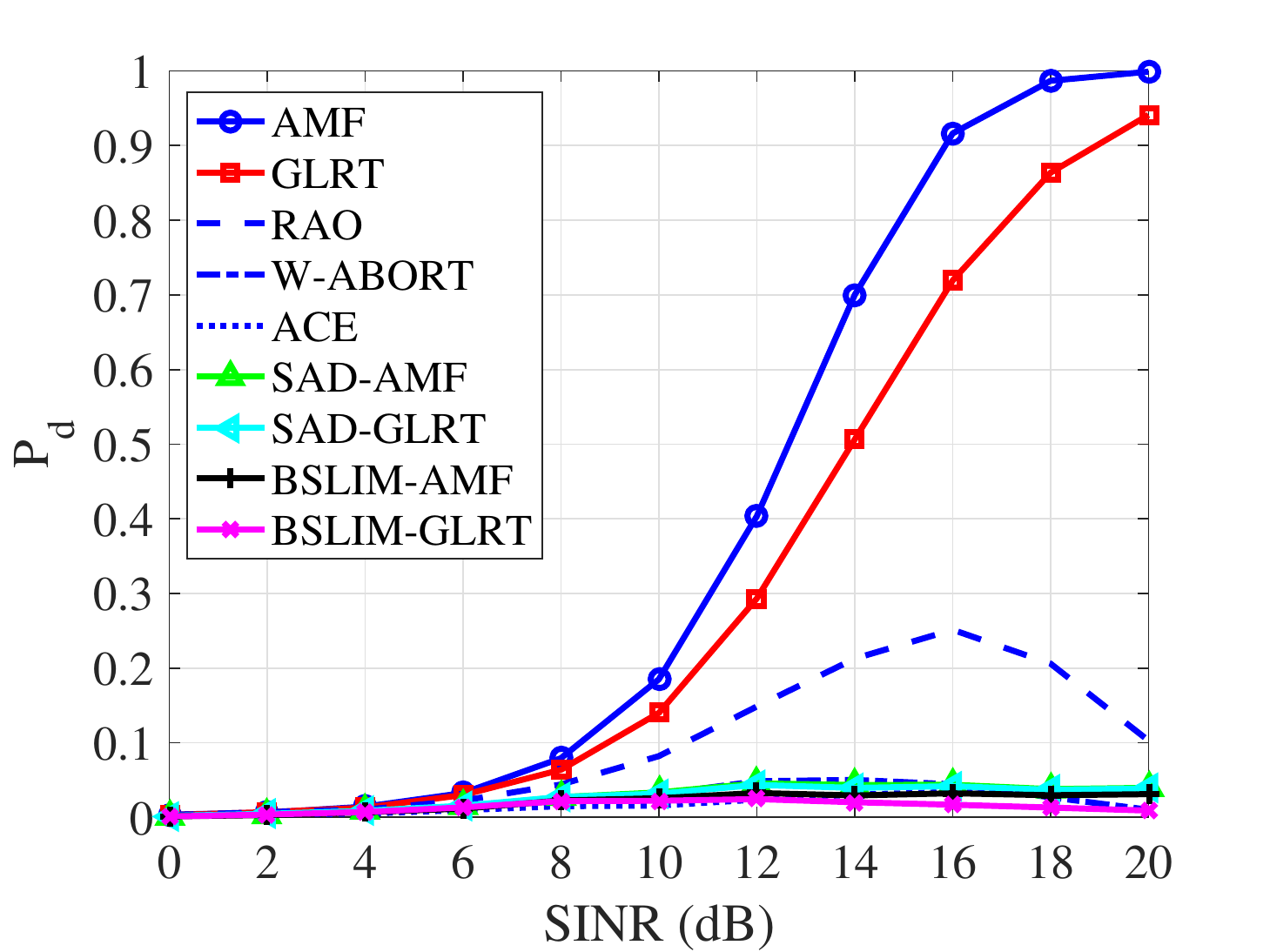}}
	\caption{$P_d$ versus SINR for $N=8$, $K=32$, $\Delta\theta=2^{\circ}$, and two $\theta_t$ values. (a) $\theta_t=0.5^{\circ}$. (b) $\theta_t=2^{\circ}$.}
	\label{mismatched_Pd_SNR}
\end{figure}
The last part of the analysis focuses on the case of mismatched signals, where the actual angle of the coherent
component is not aligned with the (nominal) angle under test. 
The next numerical examples are aimed at showing that the proposed architectures can ensure an 
enhanced selectivity. As in the case of matched targets, the influence of $K$ and SINR is assessed. 
Specifically, in Fig. \ref{mismatched_Pd_K}, $P_d$ versus $K$ is plotted for $N=8$, $\Delta\theta=2^{\circ}$, $\text{SINR}=14$ dB and two values of $\theta_t$: (a) $\theta_t=0.5^{\circ}$ (the angle of the target is not aligned with any value of $\theta_m,\;m=1,\cdots,M$, where $\theta_m$ is the $m$th angular position among the discretized antenna beam); (b) $\theta_t=2^{\circ}$ (the angle of the target coincides with a specific value of $\theta_m$ but is not equal to the nominal angle). The plots highlight that for the chosen parameters, 
the SAD-AMF, SAD-GLRT, BSLIM-AMF, and BSLIM-GLRT significantly outperform the AMF and Kelly's GLRT 
in terms of rejecting mismatched signals for a mismatch of $2^{\circ}$. 
Moreover, the $P_d$ difference between the proposed four decision schemes and the AMF and Kelly's GLRT 
grows as $K$ increases for case (b). 
Precisely, for case (b), the $P_d$ value of the SAD-AMF, SAD-GLRT, BSLIM-AMF and BSLIM-GLRT is always smaller than 0.1 for all the values of $K$. 
In contrast, the $P_d$ values of the AMF and Kelly's GLRT are 0.32 and 0.15 respectively when $K=16$, whereas 0.77 and 0.6 respectively when $K=40$. 

The improved capability of rejecting mismatched signals for the proposed four detectors is further confirmed 
in Fig. \ref{mismatched_Pd_SNR}, where $P_d$ versus SINR is plotted for $N=8$, $K=32$, $\Delta\theta=2^{\circ}$, and the same values of $\theta_t$ as in Fig. \ref{mismatched_Pd_K}. The plots highlight 
that for the AMF and Kelly's GLRT, the $P_d$ value increases as SINR increases. On the other hand, 
for SAD-AMF, SAD-GLRT, BSLIM-AMF, and BSLIM-GLRT, the $P_d$ value increases as SINR increases for case (a) whereas shows small changes for case (b). 
This is due to the fact that for case (b), the estimated target angular 
position (i.e., the non-zero element in $\hat{\alpha}$) exploiting the user parameter 
free BSLIM procedure is with high probability not aligned with the nominal angle.
\begin{figure}[htb] \centering
	\subfigure[]{\includegraphics[width=0.49\columnwidth]{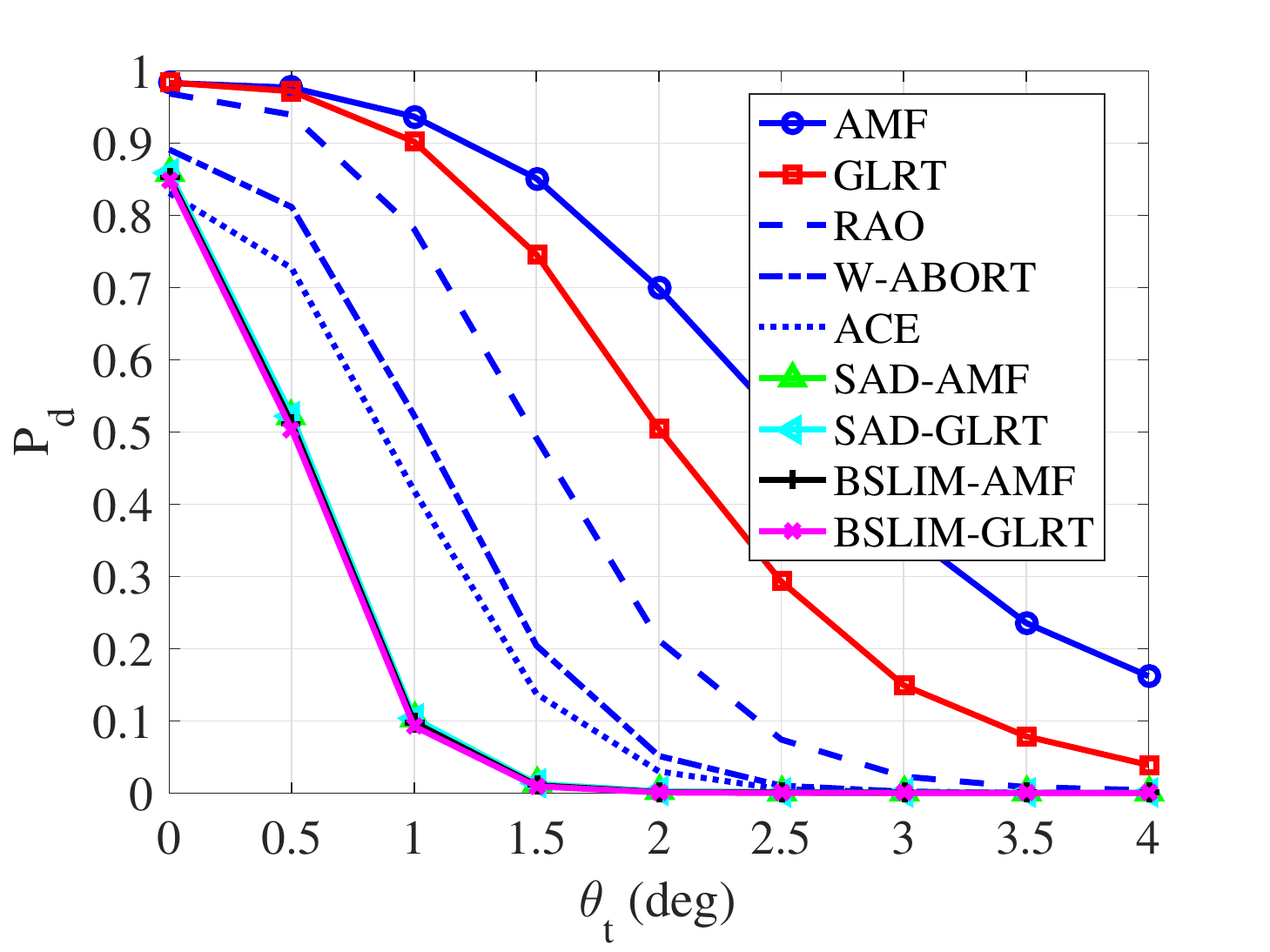}}
	\subfigure[]{\includegraphics[width=0.49\columnwidth]{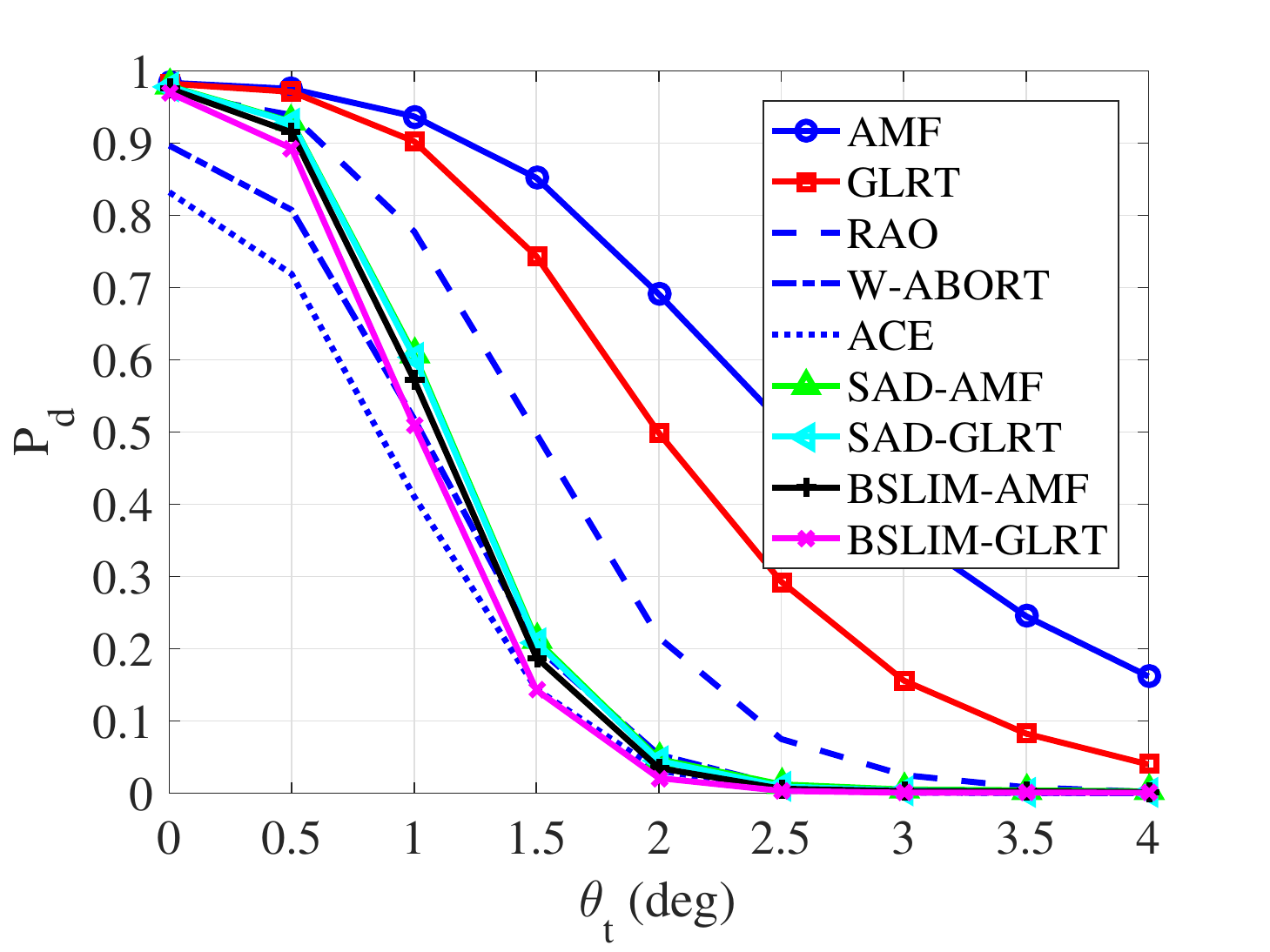}}
	\subfigure[]{\includegraphics[width=0.49\columnwidth]{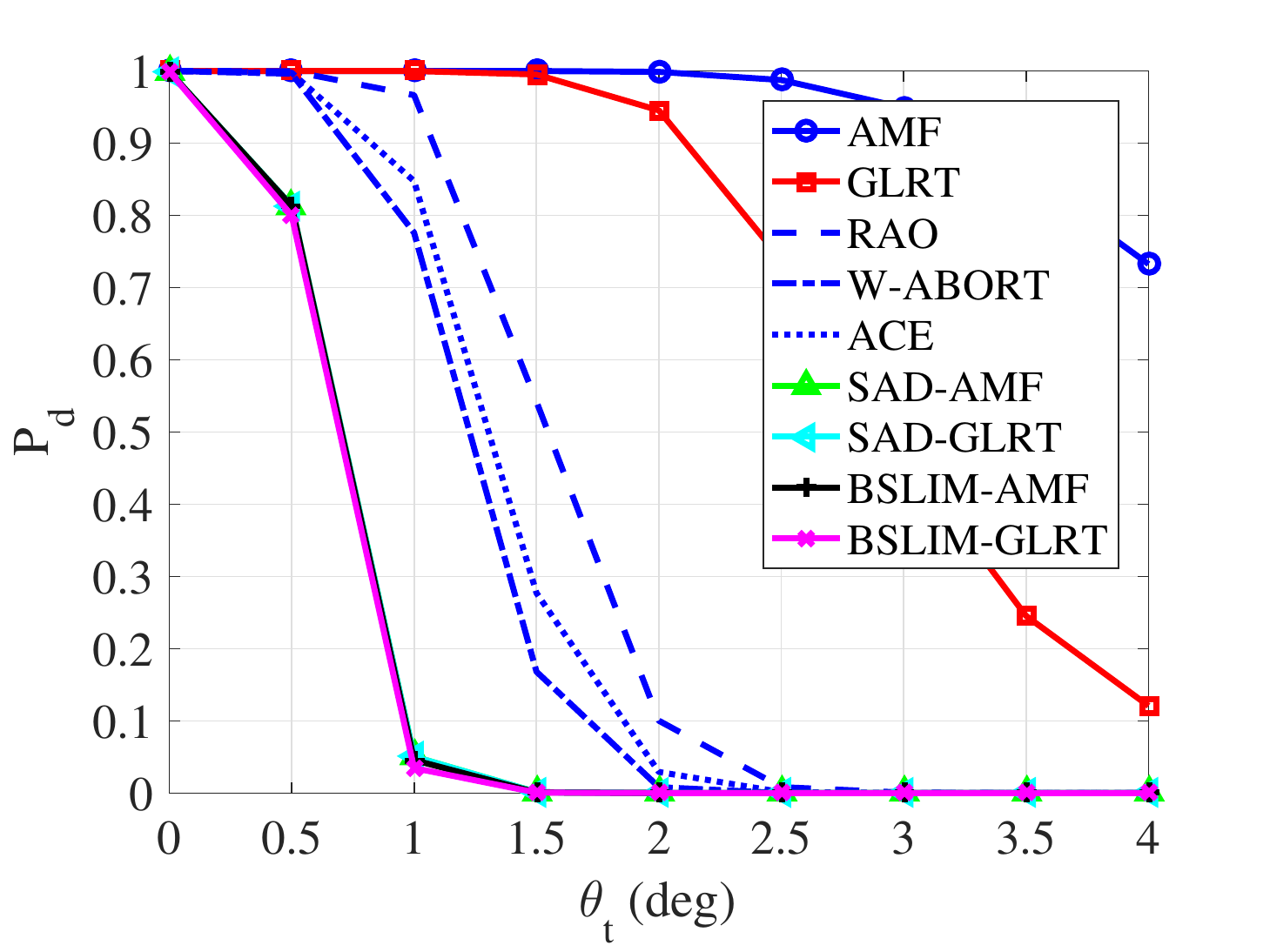}}
	\subfigure[]{\includegraphics[width=0.49\columnwidth]{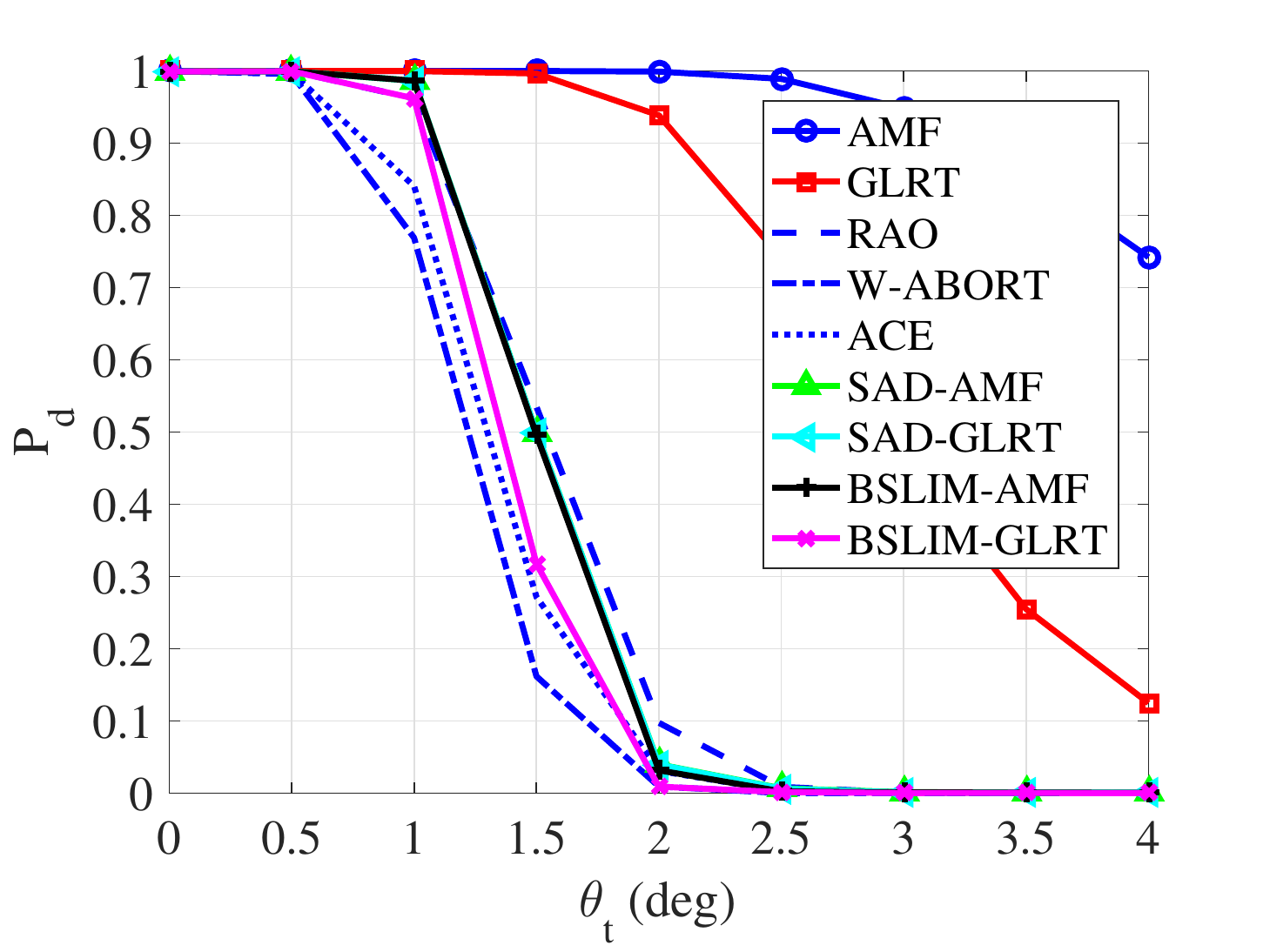}}
	\caption{$P_d$ versus $\theta_t$ for $N=8$ and $K=32$. 
	(a) $\Delta\theta=1^{\circ}$ and $\text{SINR}=14$ dB. 
	(b) $\Delta\theta=2^{\circ}$ and $\text{SINR}=14$ dB.
	(c) $\Delta\theta=1^{\circ}$ and $\text{SINR}=20$ dB.
	(d) $\Delta\theta=2^{\circ}$ and $\text{SINR}=20$ dB.}
	\label{mismatched_Pd_angle_8}
\end{figure}
\begin{figure}[htb] \centering
	\subfigure[]{\includegraphics[width=0.49\columnwidth]{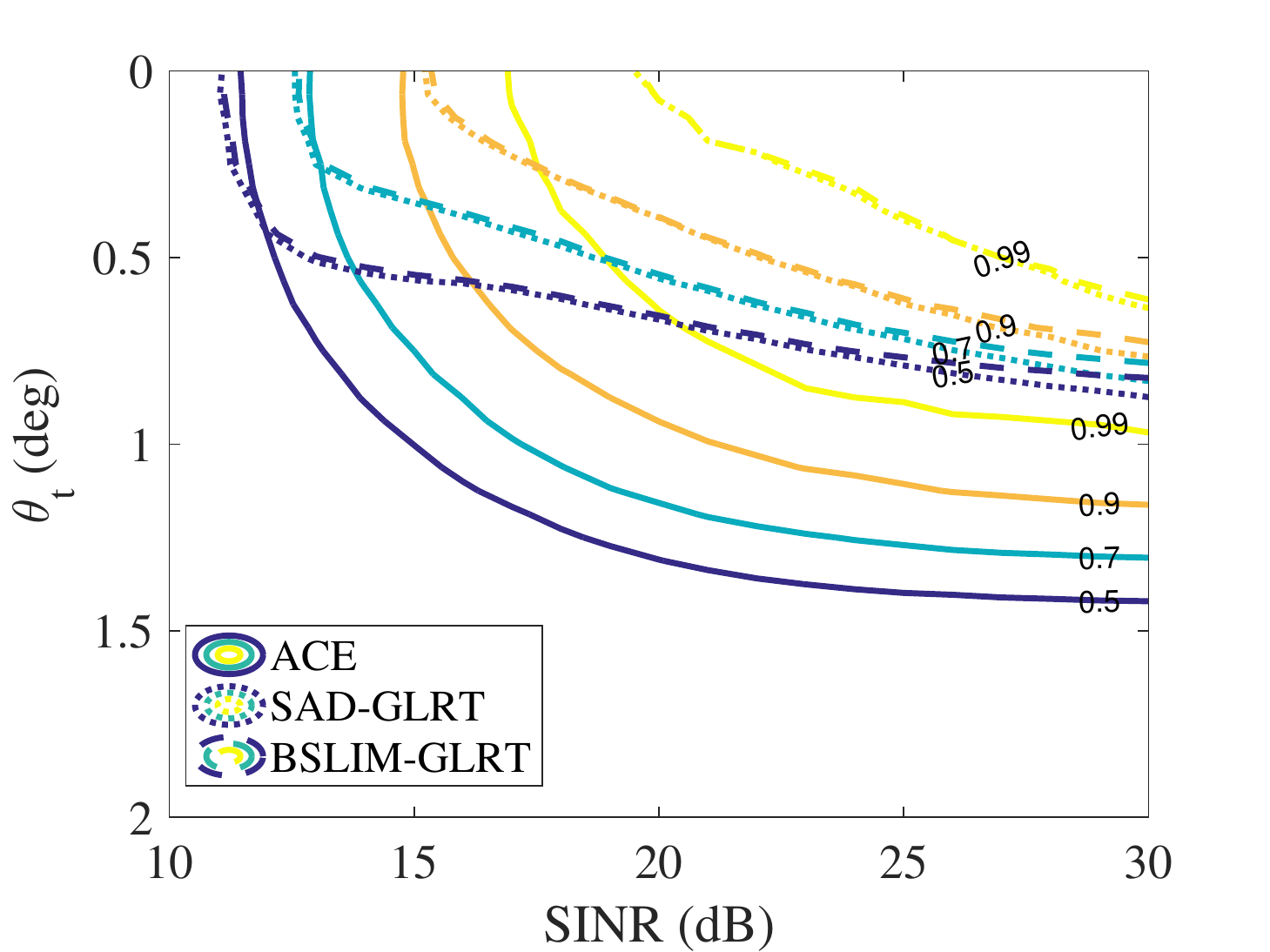}}
	\subfigure[]{\includegraphics[width=0.49\columnwidth]{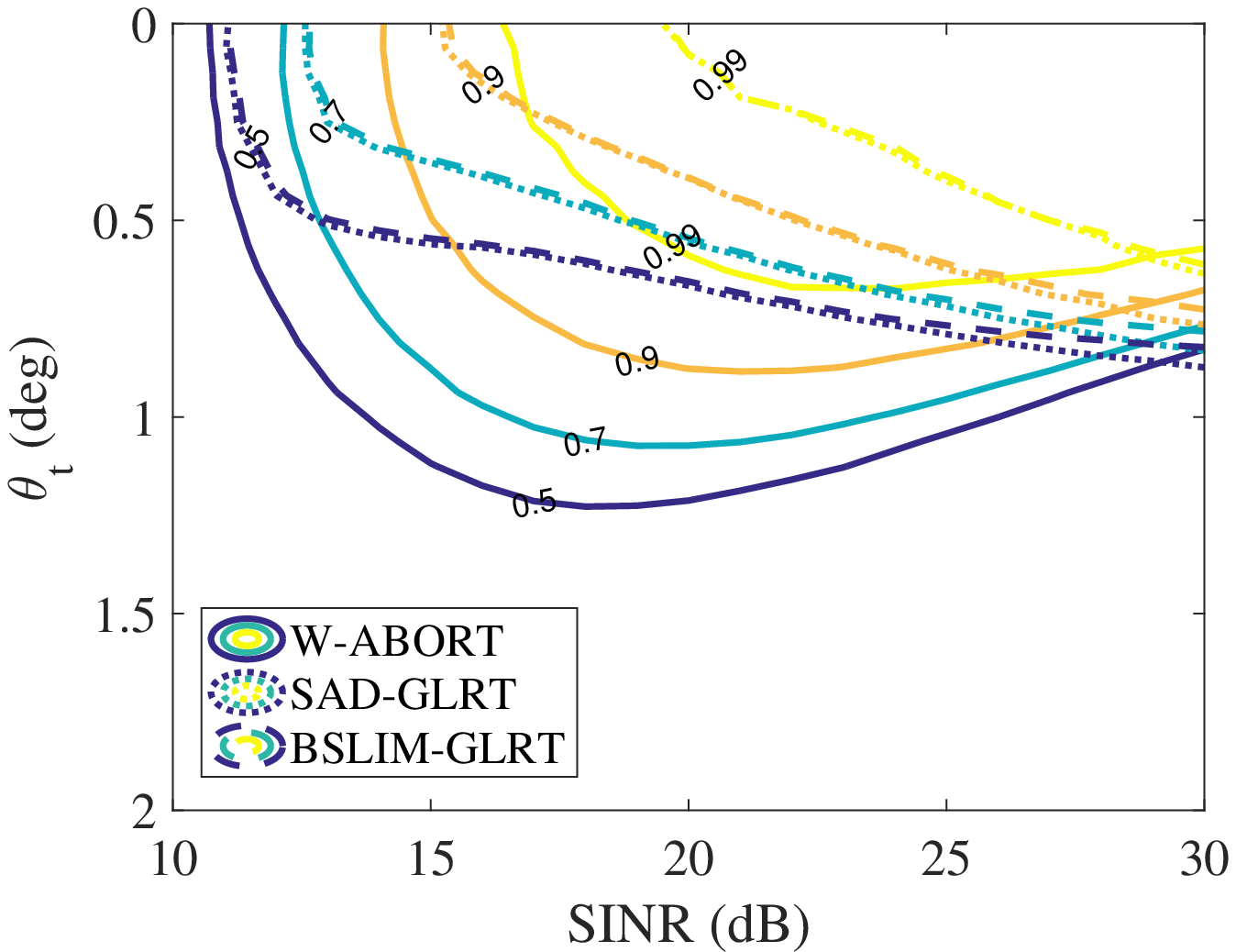}}
	\caption{Contours of constant $P_d$ for $N=8$, $K=32$, and $\Delta\theta=1^{\circ}$.
	(a) ACE, SAD-GLRT and BSLIM-GLRT. (b) W-ABORT, SAD-GLRT and BSLIM-GLRT.}
	\label{mesa1}
\end{figure}
\begin{figure}[htb] \centering
	\subfigure[]{\includegraphics[width=0.49\columnwidth]{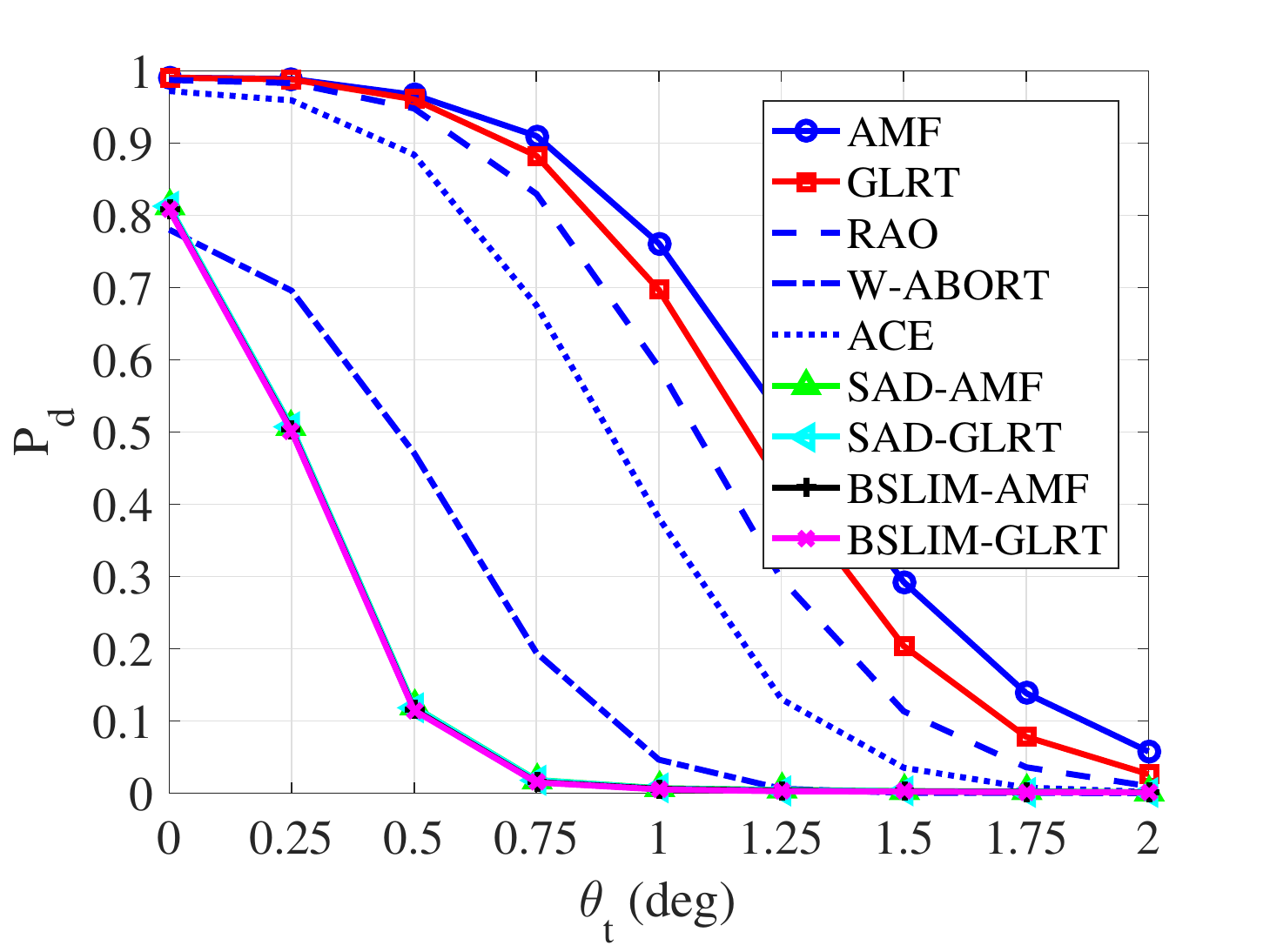}}
	\subfigure[]{\includegraphics[width=0.49\columnwidth]{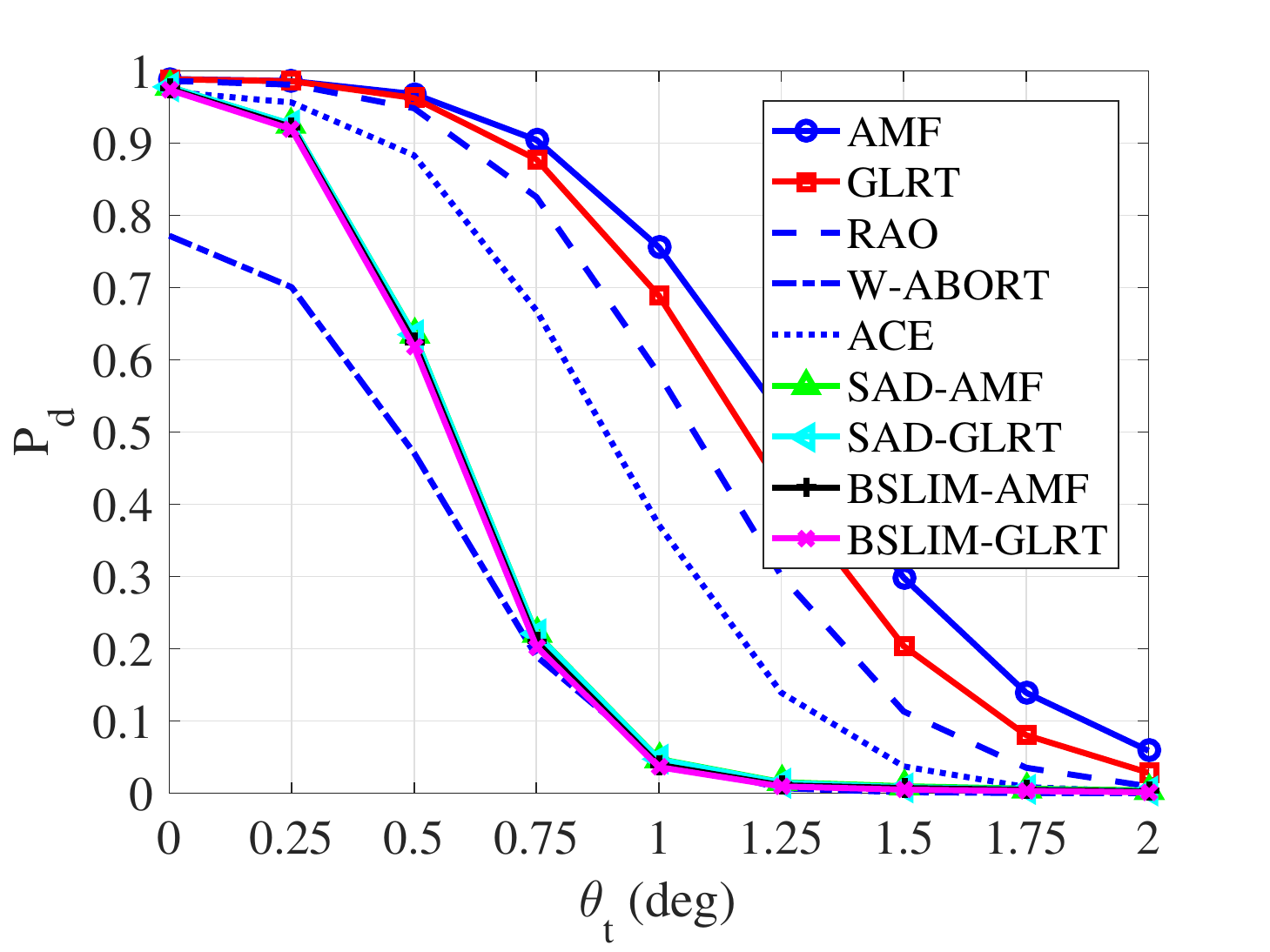}}
	\caption{$P_d$ versus $\theta_t$ for $N=24$, $K=96$, $\text{SINR}=14$ dB and two values of $\Delta\theta$. (a) $\Delta\theta=0.5^{\circ}$. (b) $\Delta\theta=1^{\circ}$.}
	\label{mismatched_Pd_angle_24}
\end{figure}

Fig. \ref{mismatched_Pd_angle_8} provides the $P_d$ curves as functions of $\theta_t$, namely
the actual azimuthal angle $\theta_t$ of the coherent component, assuming $N=8$, $K=32$, two values 
of SINR and $\Delta\theta$.
The curves show that the ACE and W-ABORT outperform the AMF, Kelly's GLRT and RAO in terms of selectivity,
whereas the new decision schemes can provide an excellent capability of rejecting undesired signals
by properly selecting the value of $\Delta\theta$ as shown in subfigures (a) and (c).
The mismatched signal detection performance of SAD-GLRT and BSLIM-GLRT in comparison with those of the ACE and the W-ABORT
are also analyzed in Fig. \ref{mesa1} for $\Delta\theta=1^\circ$, 
wherein the contours of constant $P_d$ are represented as a function of the mismatch angle, plotted vertically, 
and the SINR, plotted horizontally\footnote{Similar $2$-dimensional plots were introduced in \cite{pulsone2001about}, 
where they are referred to as mesa plots.}. It is noted that the SAD-AMF and BSLIM-AMF almost exhibit the same performance as the SAD-GLRT and BSLIM-GLRT, whose curves are not shown here for clarity. The figure highlights that the SAD-GLRT, BSLIM-GLRT, ACE, and W-ABORT share approximately the same performance for $\theta_t=0^\circ$, but the contours of SAD-GLRT and BSLIM-GLRT are more compressed towards zero with respect to those of the ACE and the W-ABORT. The quid pro quo for this enhanced selectivity is a performance degradation for matched signals
with respect to Kelly's GLRT, the AMF, and RAO as shown in Fig. \ref{figure_Pd_SNR}.
Nevertheless, the other selective architectures share more or less the same matched detection
performance as the new decision schemes. 
Moreover, increasing $\Delta\theta$ reduces the selectivity but restores the matched 
detection performance. As a matter of fact, for $\Delta\theta=2^\circ$ (subfigures (b) and (d) in Fig. \ref{mismatched_Pd_angle_8}), the new
architectures are slightly less selective than the ACE and the W-ABORT, but the latter
experience a nonnegligible loss in matched detection performance with respect to the former as
corroborated by Fig. \ref{figure_Pd_SNR}.
This behavior is further confirmed in Fig. \ref{mismatched_Pd_angle_24}, where 
$P_d$ versus $\theta_t$ is plotted for $N=24$, $K=96$, $\text{SINR}=14$ dB and two values of $\Delta\theta$. 
Comparing Fig. \ref{mismatched_Pd_angle_8} and Fig. \ref{mismatched_Pd_angle_24}, 
it is clear that for a given SINR, the capability of distinguishing targets whose angles are close each other
improves as $N$ increases.

Summarizing, a suitable selection of $\Delta\theta$ for the SAD-AMF, SAD-GLRT, BSLIM-AMF and BSLIM-GLRT 
allows for a tradeoff between the detection of matched targets and rejection of undesired signals.

\section{Conclusion}
\label{conclusion}
In this paper, we have considered the design of tunable detectors based on sparse recovery techniques in 
order to achieve an enhanced selectivity assuming colored Gaussian interference with unknown covariance matrix. 
In this context, we have introduced a user parameter free sparse recovery algorithm inspired by the SLIM method 
\cite{stoicaSparse}. Then, the estimates provided by the latter procedure have been used to devise two classes of
detectors resorting to either the two-stage detection theory (leading to the SAD-AMF and the SAD-GLRT) or
heuristic modifications of the GLRT (giving rise to the BSLIM-AMF and the BSLIM-GLRT).
Moreover, we have proved that the new decision architectures exhibit a bounded-CFAR behavior. 
Simulation results have validated that the new detectors can outperform the conventional decision schemes 
in terms of selectivity by properly choosing the value of the tuning parameter $\Delta\theta$. 
Possible future research directions might concern the extensions of the proposed framework to range-spread 
targets as well as to non-Gaussian interference. Finally, it would be interesting to test the proposed 
detectors also on real recorded data.

\section*{Acknowledgment}
The authors would like to thank Prof. A. De Maio for the exciting and interesting discussions.

\bibliographystyle{IEEEtran}
\bibliography{group_bib}

\begin{biography}[{\includegraphics[width=1in,height=1.25in,clip,keepaspectratio]{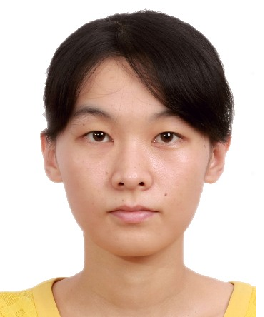}}]{Sudan Han} received the B.S. degree in information engineering, the M.S. and Ph.D. degrees in information and communication engineering, from the National University of Defense Technology, Changsha, China, in 2012, 2015 and 2019, respectively. She is currently an Assistant Researcher with the National Innovation Institute of Defense Technology, Beijing, China. Her main research interests include clutter suppression and target detection.
\end{biography}

\begin{biography}[{\includegraphics[width=1in,height=1.25in,clip,keepaspectratio]{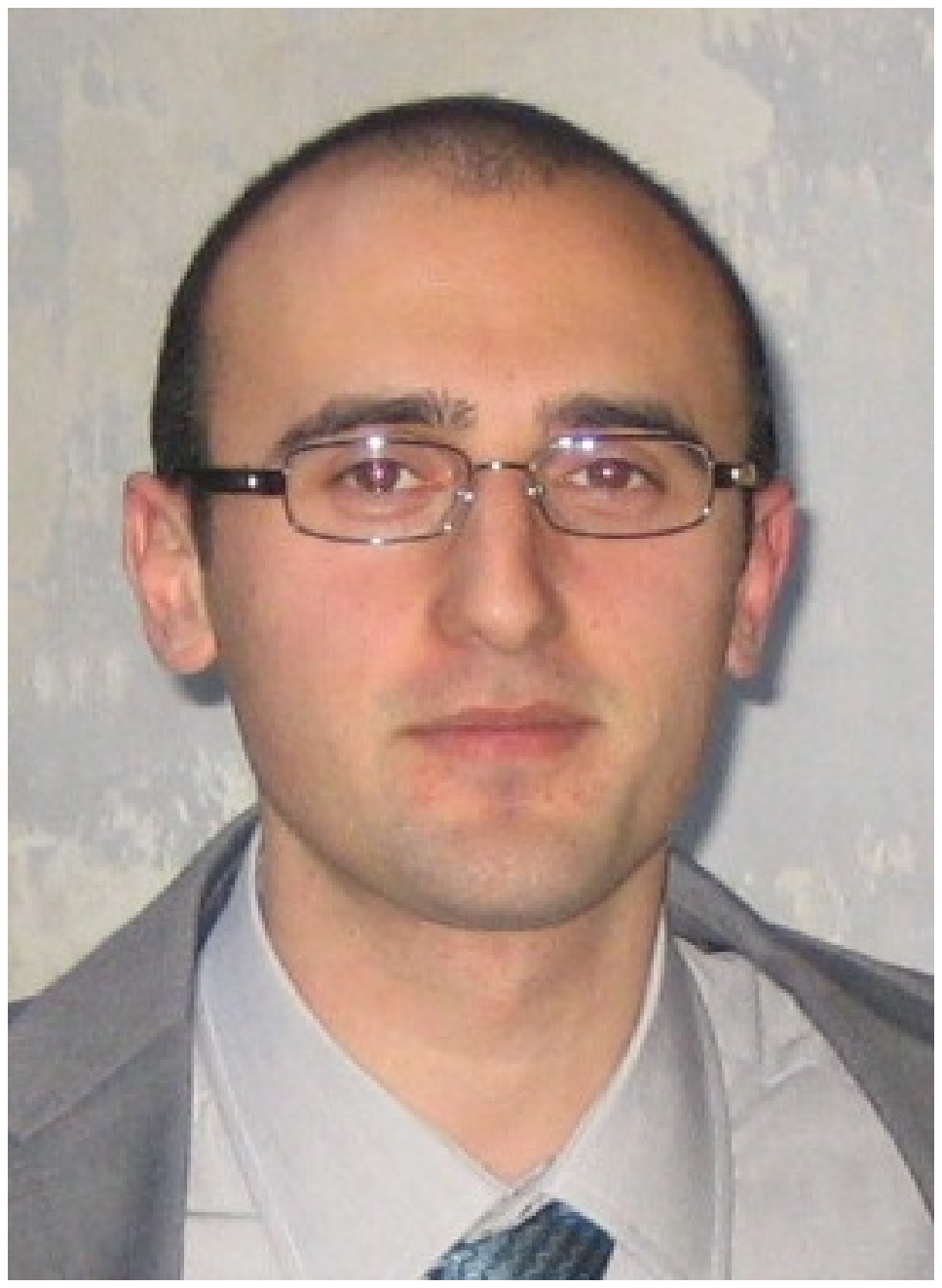}}]{Luca Pallotta}(S'12, M'15, SM'18) received the Laurea Specialistica degree (cum laude) in telecommunication engineering in 2009 from the University of Sannio, Benevento, Italy, and the Ph.D. degree in electronic and telecommunication engineering in 2014 from the University of Naples Federico II, Naples, Italy. He is currently an Assistant Professor at University of Roma Tre, Italy. His research interest lies in the field of statistical signal processing, with emphasis on radar/SAR signal processing, radar targets classification, polarimetric radar/SAR. Dr. Pallotta won the Student Paper Competition at the IEEE Radar Conference 2013.
\end{biography}

\begin{biography}[{\includegraphics[width=1in,height=1.25in,clip,keepaspectratio]{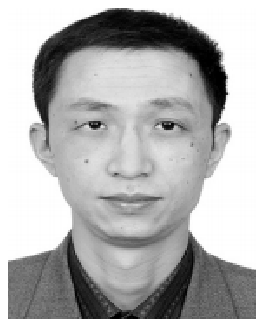}}]{Xiaotao Huang} (M'02)  received the B.S. and Ph.D. degrees in information and communication engineering from the National University of Defense Technology, Changsha, China, in 1990 and 1999, respectively.
	
He is currently a Professor with the National University of Defense Technology. His fields of interest include radar theory, signal processing, and radio frequency signal suppression.
\end{biography}

\begin{biography}[{\includegraphics[width=1in,height=1.25in,clip,keepaspectratio]{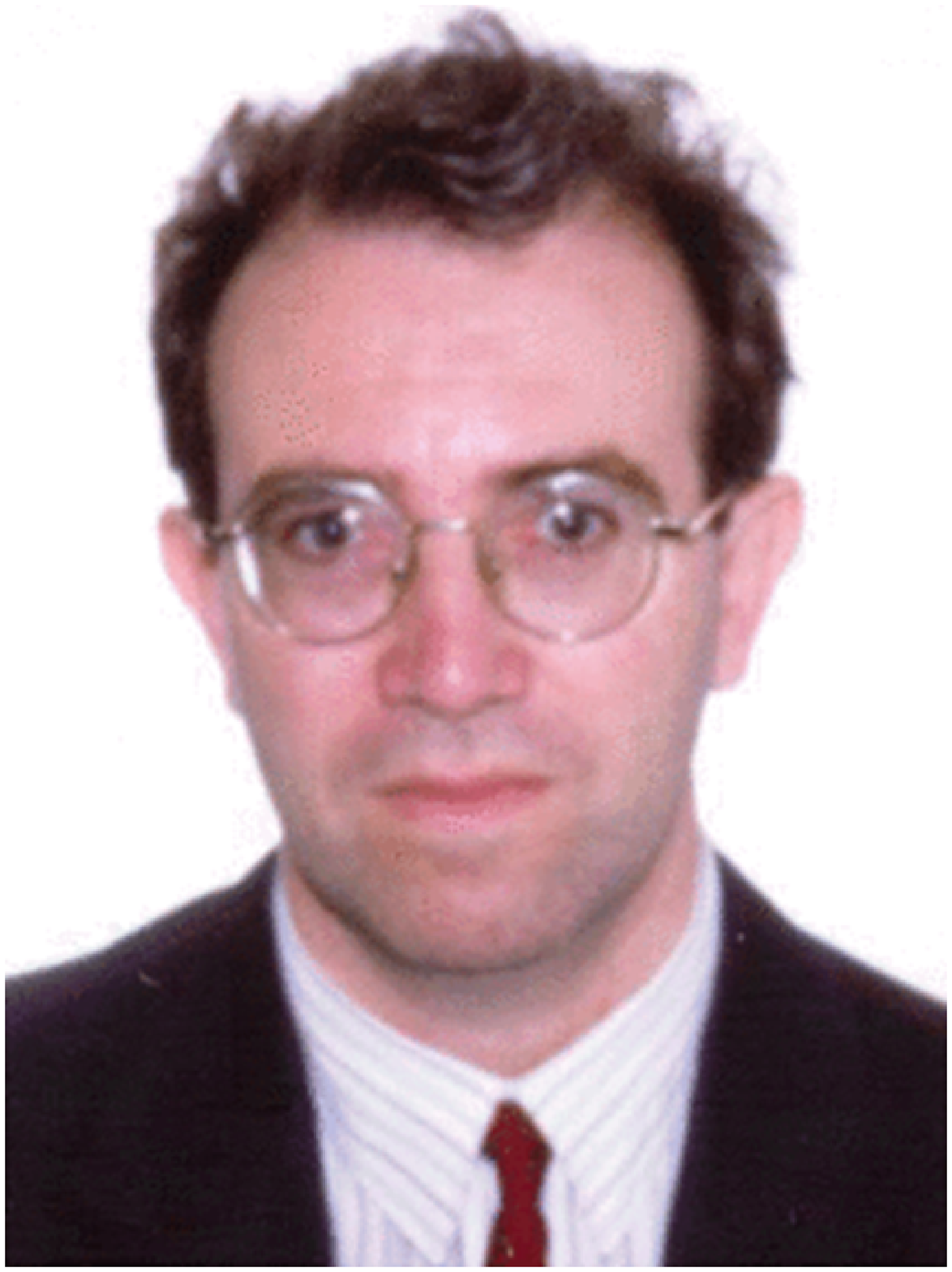}}]{Gaetano Giunta} (M'88, SM'12) received the Electronic Engineering degree from the University of Pisa, Italy, and the Ph.D. degree in Information and Communication Engineering from the University of Rome La Sapienza, Italy, in 1985 and 1990, respectively. He was also (since 1989) a Research Fellow of the Signal Processing Laboratory (LTS) at EPFL, Lausanne, Switzerland. In 1992, he became an Assistant Professor with the INFO-COM Department, University of Rome La Sapienza. From 2001 to 2005, he was with the Third University of Rome as an Associate Professor. Since 2005, he has been a Full Professor of Telecommunications with the same University. His research interests include signal processing for mobile communications, image communications and security. Prof. Giunta has been a member of the IEEE Societies of Communications, Signal Processing, and Vehicular Technology. He has also served as a reviewer for several IEEE transactions, IET (formerly IEE) proceedings, and EURASIP journals, and a TPC member for several international conferences and symposia in the same fields.
\end{biography}

\begin{biography}[{\includegraphics[width=1in,height=1.25in,clip,keepaspectratio]{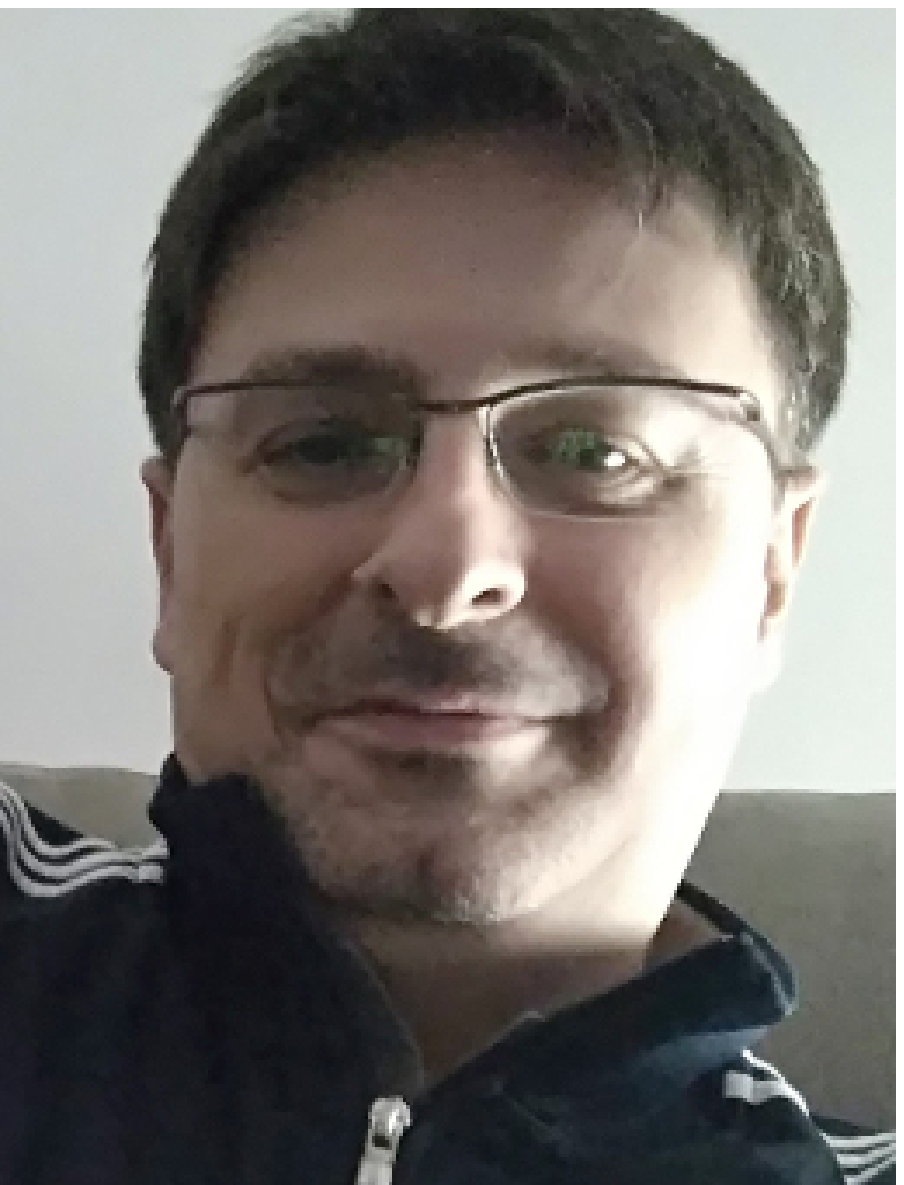}}]{Danilo Orlando} (SM' 13) was born in Gagliano del Capo, Italy, on August 9, 1978. He received the Dr. Eng. Degree (with honors) in computer engineering and the
Ph.D. degree (with maximum score) in information engineering, both from the
University of Salento (formerly University of Lecce), Italy, in 2004 and
2008, respectively. From July 2007 to July 2010, he has worked with the
University of Cassino (Italy), engaged in a research project on algorithms
for track-before-detect of multiple targets in uncertain scenarios.
From September to November 2009, he has been visiting scientist at the NATO Undersea
Research Centre (NURC), La Spezia (Italy). From
September 2011 to April 2015, he has worked at Elettronica SpA engaged as system analyst in the
field of Electronic Warfare. 
In May 2015, he joined Università degli Studi ``Niccolò Cusano'', where he is currently associate professor.
His main research interests are in the field of statistical signal processing and image processing with more emphasis on adaptive detection and tracking of multiple targets in multisensor scenarios. He has held visiting positions at the department of Avionics and Systems of ENSICA (now Institut Supérieur de l'Aéronautique et de l'Espace, ISAE), Toulouse (France) in 2007 and at Chinese Academy of Science, Beijing (China) in 2017-2019. 

He is Senior Member of IEEE; he has served IEEE Transactions on Signal Processing as Senior Area Editor and currently is Associate Editor for IEEE Open Journal on Signal Processing,
EURASIP Journal on Advances in Signal Processing, and MDPI Remote Sensing. He is also author or co-author of about 110 scientific publications 
in international journals, conferences, and books.
	
\end{biography}

\end{document}